# Robust selection of predictors and conditional outlier detection in a perturbed large-dimensional regression context

Matteo Farnè [a]* and Angelos Vouldis [b]

[a] *Department of Statistical Sciences, University of Bologna, Bologna, Italy, matteo.farne@unibo.it*

[b] *European Central Bank, Directorate General Statistics, Frankfurt am Main, Germany, angelos.vouldis@ecb.europa.eu*

**Acknowledgments**

We gratefully thank the Supervisory Statistics Division of the European Central Bank for allowing us to make research on the supervisory data of Section 4.

# Robust selection of predictors and conditional outlier detection in a perturbed large-dimensional regression context

This paper presents a fast methodology, called ROBOUT, to identify outliers in a response variable conditional on a set of linearly related predictors, retrieved from a large granular dataset. ROBOUT is shown to be effective and particularly versatile compared to existing methods in the presence of a number of data idiosyncratic features. ROBOUT is able to identify observations with outlying conditional variance when the dataset contains element-wise sparse variables, and the set of predictors contains multivariate outliers. Existing integrated methodologies like SPARSE-LTS and RLARS are systematically sub-optimal under those conditions. ROBOUT entails a robust selection stage of the statistically relevant predictors (by using a Huber or a quantile loss), the estimation of a robust regression model based on the selected predictors (by LTS, GS or MM), and a criterion to identify conditional outliers based on a robust measure of the residuals' dispersion. We conduct a comprehensive simulation study in which the different variants of the proposed algorithm are tested under an exhaustive set of different perturbation scenarios. The methodology is also applied to a granular supervisory banking dataset collected by the European Central Bank.

## Introduction

*Motivation*

Data quality is a fundamental prerequisite for any kind of quantitative analysis, and the large datasets which are becoming increasingly available present specific challenges to the task of monitoring and ensuring data quality. One critical aspect of the data quality monitoring is outlier detection, i.e. the identification of values which are either obviously mistaken or seem to be unjustified from an empirical perspective. An outlier is defined by [1] as 'an observation which deviates so much from the other observations as to arouse suspicions that it was generated by a different mechanism'. Similarly, [2] define outlier(s) as 'an observation (or subset of observations) which appears to be inconsistent with the remainder of that set of data'. Classical statistical methods, which underpin analytical tools supporting analysis of large datasets, are sensitive to the presence of outliers and, consequently, may give a distorted view of the reality underlying the data [3].

In this paper, we focus on large-dimensional datasets, where the number of variables $p$ and the number of observations $n$ are large. The datasets may also be 'fat', i.e. they may present $p > n$ or even $p \gg n$. Such datasets arise in diverse fields such as bioinformatics, economics, neuroscience, signal processing and others. For example, the cancer cell panel data used in [4] comprises 59 observations and 22,283 predictors and such dimensions are common in fMRI or gene expression studies. In economics, granular datasets are also becoming available e.g. the supermarket sales data used in [5] include 464 observations on 6,398 variables while the original banking dataset used in this paper contains 453 variables for 365 banks. Finally, in signal processing, $p$ could refer to the number of pixels which is usually significantly larger than the number of images e.g. the benchmark Eyale database [6] contains images with 480 x 640 = 307,200 pixels for 585 persons.

The detection of conditional outliers, i.e. outliers in a response variable conditional on a set of linearly related variables, presents a number of challenges [7], particularly in large dimensions. Conditional outliers may be in mean or in variance, if they present an arbitrarily large mean or variance respectively, conditional on the predictors. Recovering conditional outliers may easily become tricky in large dimensions, as the set of predictors is typically a subset, to be identified, of a high number of potential predictors. What is more, real datasets with many variables often present many zeros, multicollinearity issues, and multivariate outliers. Such idiosyncratic data features may have a negative impact on existing outlier detection methods. In addition, these features are not always known *ex ante* in real applications, and the required response time may not be sufficient to uncover them through analysis (e.g. in cases of datasets originating from diseases spreading or medical examinations). Conditional outlier detection which is effective regardless of specific statistical features is thus a challenge, and its failure may prevent the identification of hidden inconsistencies or frauds in the data.

For these reasons, we present a versatile method, called ROBOUT, for conditional outlier detection which is efficient as regards computational cost (i.e. proportional to $O(pn)$) while it works effectively for datasets featuring diverse statistical properties. Specifically, the method is robust to a number of idiosyncratic features of the dataset (such as sparsity or multivariate outliers) while it integrates a component robustly identifying the relevant set of predictors. We conduct a comprehensive simulation study including scenarios featuring leverage outliers, various levels of data sparsity, different relative data dimensions (cases of both $p > n$ and $p < n$), and existence of multicollinearity. ROBOUT is efficient and effective in this broad range of situations, thus providing a feasible and reliable answer to conditional outlier detection in large dimensions.

*State of the art*

A recent cornerstone of the outlier detection literature is [8] which reviews several outlier detection methods, including those suitable for high-dimensional datasets. Those methods are useful to identify outliers in several senses. However, some outlier detection approaches are clearly inefficient on large datasets e.g. distance-based outlier detection methods quickly become very heavy computationally. In addition, in the presence of weak correlations among the dataset variables, the curse of dimensionality in the form of the so-called 'concentration effect' [9] may take its toll, and render distance measures unsuitable to identify outliers.

Overall, multidimensional outlier detection requires strategies tailored for use in high dimensions [10-12], as the most important variables to understand the data content are often part of a variable subspace [13,14]. High-dimensional outlier detection methods include robust deep autoencoders [15], axis parallel subspaces [16], outlier scores aggregation [17], probabilistic multivariate conditional outliers [18], and feature selection in high dimensions [19].

Large datasets, however, appear more suitable for outlier detection in the regression context. As [20] notes, the large size of the data requires automated techniques for the identification of subsets of variables which are statistically linked. In such datasets, the challenge is to identify the critical predictors and then to perform a conditional outlier detection for the variables of interest. To this end, we need to define both a variable selection and a robust regression step, in order to recover consistently both the true set of predictors and the regression coefficients. This type of outlier detection may spot outliers which could remain unnoticed if single variables are considered separately, thus taking advantage of the information content of the large cross section.

Among variable selection methods, the main idea in the literature is still the LASSO [21], with its many variants (like EXTENDED [22], FUSED [23], GRAPH [24], GROUP [25], SPARSE-GROUP [26] and ADAPTIVE [27]). Unfortunately, none of those methodologies is robust, as they all employ a least square type loss. Other variable selection methods, like SURE [28], SCAD [29], the Dantzig selector [30], Forward Regression [3] and SLOPE [31] are effective for fat datasets under the condition that no outliers are present in the response variable. In contrast, two LASSO variants, namely MD-LASSO [32] and LAD-LASSO [33], are robust to the presence of outliers in the response variable as they replace the least squares loss by a robust alternative.

Once a robust variable selection is performed, we need a robust conditional outlier detection method. A good proposal, based on $l_1$ regression, to spot outliers in the principal component spaces is [34]. Another relevant work [35] performs multiple outlier detection in a high-dimensional regression context using SURE [28] and least trimmed squares [36]. The mean-shift model [37] spots conditional outliers by robustly estimating an intercept for each observation, but it is not intended to recover conditional outliers in variance. A method like Xcluster [38], which tolerates well multivariate outliers in the predictors (also called leverage outliers), is not prepared to cope with a sparse design matrix.

The main integrated methods performing both robust variable selection and conditional outlier detection that are proposed in the literature are SPARSE-LTS [4] and RLARS [43]. The former is based on the simultaneous optimization of a trimmed least squares loss and a LASSO penalty while the latter is the robustified version of LARS [44], where the covariance matrix of the features is robustly estimated via an adjusted Winsorization step. SPARSE-LTS is not robust against conditional outliers in variance, particularly when the sample size is large. This is due to the trimming step, which leads in that case to a systematically too large residual variance estimate. The consequence is the heavy masking of conditional outliers, which also occurs when the design matrix is sparse. Instead, RLARS is not even feasible in the case of severe sparsity, as the robustified covariance matrix of the features cannot be computed anymore. RLARS is also expected to be exposed to severe inconsistency when leverage outliers are present in the predictors and $p > n$.

The computational cost of RLARS is $O(p \log p \, n \, K)$, where K is the chosen number of predictors, while for SPARSE-LTS it is $O(pn^2)$, due to the cost of sorting squared residuals. Therefore, SPARSE-LTS is clearly not scalable to a large sample size, while RLARS requires to select a small number K of required predictors to work fast. SPARSE-LTS and RLARS will be the benchmark methods against which the effectiveness of our proposed method is tested.

*Contributions*

In the presence of conditional outliers in variance, when the potential predictors are many and potentially sparse and the actual predictors contain leverage outliers, providing a fast and effective answer to conditional outlier detection is not trivial. In particular, we have seen in the previous section that both variable selection and conditional outlier recovery are critically challenged. In this paper we propose ROBOUT procedure, where both these steps are designed to cope with the different perturbation scenarios without affecting the effectiveness of the outlier recovery.

First of all, it is crucial to use a variable selection method which does not depend on a single statistic like the residual variance. In this respect, an optimal candidate solution appears to be SNCD [45], a method which optimizes simultaneously a Huber loss or a quantile loss of the residuals and a LASSO penalty (henceforth, we call the two options of using a Huber or a quantile loss function SNCD-H and SNCD-Q respectively). SNCD is scalable to both large sample sizes and high dimensions, as its computational cost is $O(pn)$. For these reasons, SNCD represents the first step of our proposed ROBOUT procedure.

Differently from SPARSE-LTS, SNCD seems to be robust to conditional outliers in variance, as it is based on robust losses in that respect. When the chosen quantile is the median, as in our case, SNCD-Q is conceptually equivalent to LAD-LASSO. Therefore, SNCD-H and SNCD-Q should consistently select the relevant predictors, even if we expect SNCD-Q to be somehow more susceptible to swamping than SNCD-H when the number of potential predictors is high, because the median loss is the most robust possible (with a breakdown point of 50%). Therefore, it is critical to test the performance of both methods for different p/n ratios.

The use of either SNCD-H or SNCD-Q alone is not sufficient generally for the purposes of conditional outlier detection, as these methods do not provide a consistent estimation of the residuals and consequently of the residual scale. This is because they are M-estimators not derived by an iterative weighted least squares algorithm. For this reason, we combine the variable selection options with a robust regression method, to robustly estimate the residuals and their variance. Then, assuming normality on the rescaled residuals, an outlier detection rule can subsequently be defined.

For the robust regression step, we examine the performance of three robust regression methods in the literature, i.e. LTS [39], GS [40], and MM [41], under our perturbation scenarios. Therefore, LTS, GS and MM will be the robust regression alternatives for ROBOUT, as there is room to meaningfully explore the empirical capability of these three established methods to estimate regression coefficients from a sparse design matrix with leverage outliers and to consequently spot conditional outliers in variance in the response variable in a robust way. A limitation of LTS noted in the incumbent literature is that its effectiveness may be weak under multicollinearity [42].

The overall uncertainty of ROBOUT, formally defined in Section 2, lies in the effect of the outlined perturbed conditions on both steps (variable selection and robust regression). The systematic simulation that we undertake in Section 3 aims to identify the optimal design of the ROBOUT procedure, i.e. the optimal combination of SNCD-H or SNCD-Q in the first step, with LTS, GS or MM in the second step, placing emphasis on the versatility of the ensuing conditional outlier detection procedure under an exhaustive number of perturbation scenarios for different p/n ratios Subsequently, we apply the ROBOUT procedure to banking data collected by the European Central Bank in Section 4.

**ROBOUT: a tandem approach for conditional outlier detection**

*Data*

Our dataset consists of n numerical observations on $p + 1$ variables. Our aim is to identify conditional outliers for one specific response variable $y$ corresponding to a specific variable of the input dataset.[1] Let us define a set of outlier indices $O$ with $O \subseteq \{1, \cdots, n\}$ indicating the observations of $y$ which are outliers and $|O| = [\alpha n]$, with $\alpha \in [0, 0.5]$ indicating the degree of contamination with outliers. Let us denote by $D$ the set of predictor indices, such that $|D| = K$ and $D \subseteq \{1, \cdots, p\}$. Therefore, the response variable can be expressed in terms of the following regression model

$$y = X_D \beta + \varepsilon, \tag{1}$$

where $y$ is the $n \times 1$ vector of the response variable, $X_D$ is the $n \times K$ matrix of predictors (whose columns are indexed by $D$), $\beta$ is the $p \times 1$ vector of regression coefficients and $\varepsilon$ is the $n \times 1$ vector of residuals.

Let us consider the regression model for the single observation $i \in \{1, \cdots, n\}$:

---

[1] The procedure developed can then be applied iteratively to all the $p + 1$ variables of the dataset if the aim is to identify conditional outliers in the whole dataset.

$$y_i = \mathbf{X_{D,i}}\boldsymbol{\beta} + \varepsilon_i, \tag{2}$$

where $\mathbf{X_{D,i}}$ denotes the i-th row of $\mathbf{X_D}$. For each $i^* \in O$, we assume that the conditional outliers in variance are generated by $\varepsilon_{i^*} \sim N(0, m\sigma^2)$, with $m > 1$, while $\varepsilon_{i^*} \sim N(0, \sigma^2)$ for $i^* \notin O$. In general, we set $\mathbf{X_{D,i^*}}' \sim N(\mathbf{0}, \mathbf{I})$ for $i^* \in \{1, \cdots, n\}$. In some of the simulation scenarios we also allow for the existence of leverage outliers for which $\mathbf{X_{D,i^*}}' \sim N(\mathbf{0}, m\mathbf{I})$ for $i^* \in O$.

We call $D'$ the complementary set of $D$ with respect to the set $\{1, \cdots, p\}$. The columns of the $n \times (p - K)$ matrix $\mathbf{X_{D'}}$ contain the variables indexed by $D'$. In some applications it is expected that $\mathbf{X_{D'}}$ is element-wise sparse. We assume zeros to be randomly positioned and to be $\gamma \times n \times (p - K)$, with $\gamma \in [0,1)$ determining the degree of sparsity.

In the setting described above, it has been assumed that multicollinearity has been removed from the data. However, we will also investigate in the Appendix one scenario with combined sparsity and multicollinearity, together with a couple of basic scenarios with conditional outliers in mean, both alone and combined with leverage outliers in the predictors.

*Competitor methods*

The perturbed statistical setting that we investigate, featuring sparsity in the design matrix, presence of leverage outliers in the predictors and presence of conditional outliers in variance in the response variable, produces multiple consequences for the standard OLS regression when applied to Eq. (2). Sparsity often renders the determinant of the covariance matrix of the design matrix close to 0, which leads to the instability of estimated coefficients. Leverage outliers perturb the covariance matrix of the predictors and increase the variance of estimated coefficients. The presence of observations with an outlying residual variance also violates the constant residual variance assumption, which causes outlying points to have a dramatic impact on OLS estimates.

An alternative approach suggested in the literature is SPARSE-LTS [4], which is built as a LASSO 'trimmed' method, based on the following minimization objective:

$$Q(H, \boldsymbol{\beta}) = \sum_{i \in H} \varepsilon_i^2 + \alpha n \sum_{j=1}^{p} |\boldsymbol{\beta}_j|, \qquad (3)$$

where $H \subseteq \{1, \ldots, n\}$ is the subsample of observations with cardinality $[(1 - \alpha)n]$, $\alpha \in [0,0.5]$, producing the minimum for $Q(H, \boldsymbol{\beta})$ with respect to $\boldsymbol{\beta}$. The optimal $\boldsymbol{\beta}$ is then computed via a Sparse Fast-LTS algorithm, which performs at each step a LASSO estimation on the $100 \times (1 - \alpha)\%$ observations with the smallest squared residuals.

Such procedure suffers from similar problems as the OLS method under our perturbed conditions, particularly when n~p and both are large. In that case, observations with an outlying residual variance may be severely masked, due to the systematic underestimation of the residual variance caused by the outlying $\sigma$.

RLARS [43] follows instead another strategy: it deals with outliers by rendering them uninfluential in the iterative computation of the correlation matrix needed to retrieve predictors. We stress that RLARS does not tolerate a severe sparsity of the design matrix, as the median absolute deviation (MAD) of at least one of the features is likely to be 0, thus making the adjusted Winsorized covariance matrix unfeasible. Severe masking and swamping effects are also expected in the presence of conditional outliers and leverage outliers in the predictors as $p > n$, because the recalled robust covariance matrix is built upon the sample covariance matrix.

*Definition*

We present here our proposed conditional outlier detection method, ROBOUT, consisting of three main steps, namely, variable selection, robust regression and outlier detection.

*Variable selection*

Starting with our target response variable $y_i$, $i = 1, \cdots, n$, we aim here to select a relevant set of predictors, considering the possibility of zero entries and multivariate outliers in the matrix of potential predictors **X**. Consequently, we aim to identify a model such as Eq. (2). One option that we consider is to use the SNCD-H objective function:

$$\min_{\{\beta_j\}_{j=0,1,\ldots,p}} \sum_{i=1}^{n} \left( \rho_H(\varepsilon_i) + \lambda \sum_{j=1}^{p} |\beta_j| \right), \tag{4}$$

where $\varepsilon_i = y_i - \beta_0 - \sum_{j=1}^{p} \beta_j X_{ij}$ and $\rho_H(t) = \frac{1}{2}\{|t| + (2\tau - 1)t^2\}$, $t \in R$, is the Huber weight function. Eq. (4) estimates an elastic-net penalized Huber loss regression, optimized by using the semi-smooth Newton coordinate descent algorithm presented in [44]. Weighting observations renders the results robust in the face of the perturbed conditions, because it annihilates the influence of conditional outliers. The Huber penalty works in fact under the assumption that outliers come from a generic contaminating distribution, not based on any specific statistic like the residual variance [5].

The other robust alternative that we consider, namely SNCD-Q, is like (4) substituting $\rho_H(\varepsilon_i)$ with $\rho_Q(\varepsilon_i)$, where $\rho_Q(t) = t\left\{\frac{1}{2} - I(t < 0)\right\}$, $t \in R$, replaces the Huber penalty $\rho_H(\varepsilon_i)$. SNCD-Q estimates an elastic-net penalized median loss regression, optimized in the same way.

*Robust regression*

While SNCD provides robust estimates of regression coefficients, no robust estimate of the residual scale is provided. This happens because the objective function is not optimized via a reweighted least squares method, but deriving KKT conditions. At the same time, we can expect that SNCD is consistently retrieving the predictors under our perturbed conditions. What is more, its computational cost is both linear in p and n. Therefore, we subsequently apply a robust regression method on the recovered predictors, to estimate consistently the residual scale. We test the following three methods:

- least trimmed squares (LTS) estimation, which identifies the $100 \times (1 - \alpha)\%$ most concentrated observations and estimates the model on those via OLS;
- MM-estimator, which is a 3-stage procedure, based on M-estimation, ensuring at a given breakdown point much higher efficiency than LTS. The estimator minimizes a Huber function of the residuals and uses a robust initialization of the coefficients β and the residual scale σ;
- GS-estimator, which estimates the residual scale minimizing a Tukey biweight function of scaled residuals. The method is based on the optimization of a generalized function of the residual scale, guaranteeing minimum max-bias.

Under our perturbed conditions, LTS is supposed to suffer from a masking effect, as points with outlying variances increase the estimated residual variance too much, and the same does the presence of leverage outliers in the predictors. MM should be more robust than LTS in both cases, but also more prone to swamping effects, as it tends to underestimate the estimation variability. GS is computed by estimating the residual variance via the residual differences, and therefore it should also be more susceptible to masking, particularly in presence of conditional outliers in variance.

Concerning computational cost, we know that both LTS and MM share a burden of $O(n)$ operations. This is due to the use of Fast-LTS [46] for LTS, and to Fast-S [47] and the Design Adaptive Scale [48] for MM. GS has instead a cost proportional to $O(n^2)$.

*Outlier detection*

A robust calculation of the residuals' dispersion should exclude the impact of outliers i.e. it should be conducted by assigning zero weights to the observations which are deemed to be outliers from an initial dispersion estimate.

Specifically, for the LTS method the initial dispersion estimate $\hat{\sigma}_0$ is calculated as the MAD of estimated residuals: $\hat{\sigma}_0 = \sqrt{\text{median}_i(\hat{\varepsilon}_i^2)}/\phi^{-1}(0.75)$, where $\hat{\varepsilon}_i = y_i - \hat{\beta}_0 - \sum_{j=1}^{p} \hat{\beta}_j X_{ij}$ is the estimated residual for each i in $\{1, \ldots, n\}$ and $\phi$ is the standard normal distribution function. A null weight ($w_i=0$) is assigned to the $[\alpha n]$ points with the largest absolute scaled residual $\hat{\varepsilon}_i/\hat{\sigma}_0$, $i \in \{1, \cdots, n\}$, otherwise it is assigned $w_i=1$. The reweighted residual variance is calculated as $\hat{\sigma} = \sqrt{\frac{\sum_{i=1}^{n} w_i \hat{\varepsilon}_i^2}{\sum_{i=1}^{n} w_i}}$, and then multiplied by a finite sample correction and a consistency factor (see [50]) to get the final scale estimate.

Concerning the MM method, the robustly estimated scale $\hat{\sigma}$ is obtained by the Fast-S algorithm. Concerning the GS method, robust Mahalanobis distances of the residuals are computed and observations exceeding $\sqrt{\chi^2_{L,1-\alpha/2}}$ (where $\alpha = 0.05$) are assigned a null weight $w_i$. The residual scale is then robustly estimated accordingly.

SPARSE-LTS identifies regression outliers and then provides final estimates of coefficients and residual scale in a way which is specular to that of the LTS. RLARS also uses the same procedure, but applied to the last fit of the calculated sequence of optimal sub-models.

At the end of the process, the points with rescaled residuals $\hat{\varepsilon}_i/\hat{\sigma}$ larger than $\phi^{-1}(0.995)$ in absolute value are flagged as conditional outliers.

Consequently, six variants of our proposed methodology can be formulated. In the first step either SNCD-H or SNCD-Q can be used while in the second step there are three options, LTS, MM or GS. Our simulation study aims both to compare our proposed approach to competitor methods but also to identify the optimal design of our general method with respect to its constituent components under diverse settings and also considering the overall performance of each design.

**A comparative simulation study**

*Simulation settings*

In this section, we describe a simulation study which compares the performance of our ROBOUT procedure (with its six variants) and its direct competitors i.e. SPARSE-LTS and RLARS. We consider four macro-scenarios with conditional outliers in variance, namely:

(1) model (2) without leverage outliers and sparsity proportion $\gamma=0$;
(2) model (2) without leverage outliers and sparsity proportion $\gamma=0.3$;
(3) model (2) with leverage outliers and sparsity proportion $\gamma=0$;

(4) model (2) with leverage outliers and sparsity proportion γ=0.3.

We also examine different relative dimensions for each of the four scenarios and assume three different combinations of p and n:

    a. $p = 100$, $n = 200$ (i.e. $p < n$);

    b. $p = 200$, $n = 100$ (i.e. $p > n$);

    c. $p = 500$, $n = 50$ (i.e. $p \gg n$).

All scenarios are simulated with $m = 3,5,\cdots,19$, $\alpha = 0.1$, $\beta_0 = 10$, $\beta_k \sim U(5,15)$ or (alternatively) $\beta_k \sim U(-15,-5)$, with $k = 1,\cdots,K$, and $K = 3$.

The aim of the simulation study is to test the relative performance of the six variants of the ROBOUT procedure: SNCD-H+LTS, SNCD-H+GS, SNCD-H+MM, SNCD-Q+LTS, SNCD-Q+GS, SNCD-Q+MM, in presence of conditional outliers in variance, leverage outliers in the predictors, and sparsity in the design matrix. In addition, we measure the effectiveness of RLARS and SPARSE-LTS under the same conditions. The perturbation parameter m plays a crucial role to rank the performance of the different methods, which may be measured both in terms of masking and swamping.

For each $m = 3,5,\cdots,19$ and each of the eight methods, we use the following measures of outlier detection performance:

- the masking rate MR, defined as the proportion of masked outliers over the true number of outliers;

- the swamping rate SR, defined as the proportion of swamped outliers over the number of recovered outliers;

- the F1-score, which is an overall performance measure balancing masking and swamping, defined as $\frac{2(\text{PREC} \times \text{REC})}{\text{PREC}+\text{REC}}$, where the precision is equal to $\text{PREC} = 1 - \text{MR}$, and the recall is equal to $\text{REC} = 1 - \text{SR}$.

All measures range from 0 to 1.

To assess the performance of the variable selection step, we calculate:

- the mean number of masked predictors, MP;

- the mean number of swamped predictors, SP;

- the average between MP and SP: $\text{AP} = \frac{\text{MP}+\text{SP}}{2}$.

We also provide two R functions, 'cond_out_gen' and 'robout', performing respectively the conditional outlier generation and the ROBOUT procedure with all possible options. 'cond_out_gen' also includes the option to generate conditional outliers in mean, or to impose a fixed correlation among the features, or to annihilate a fixed proportion of rows in the data matrix. 'robout' computes all six possible options of the proposed procedure.

*Simulation results*

In Figure 1, we report the outlier detection measures for scenario 1a. Under this scenario, the predictors are perfectly recovered by all the methods (SNCD-H, SNCD-Q, SPARSE-LTS and RLARS). Therefore, the SNCD-H and SNCD-Q options of ROBOUT work almost equally well for each robust regression method. In this $p < n$ context, LTS is the best ROBOUT option, followed by SPARSE-LTS and the MM options of ROBOUT, which are subject to swamping effects. On the contrary, the performance of the GS options of ROBOUT and RLARS are quite worse: GS presents both masking and swamping rates around 10%, while RLARS presents a masking rate over 30% for $m = 3$ (slowly decreasing as m increases).

[Figure 1 here]

In Figure 2, the results for scenario 1b are displayed. In such $p > n$ case, the recovery of predictors is still very good by almost all methods, with only SNCD-H and SNCD-Q showing a slight deterioration of their performance due to masking and swamping effects (with a mean of 0.1 out of 3 predictors). Concerning outlier detection, we can see that the masking rate of all methods is very high for small values of m. As $m > 10$, instead, the MM options of ROBOUT perform acceptably, with a masking rate under 20% and a swamping rate under 10%. At the same time, the LTS options stand around a 25% masking rate and the GS options around a 30% masking rate. SPARSE-LTS is generally affected by a severe masking (apart from very large values of m), while RLARS fails completely.

[Figure 2 here]

In Figure 3, we can observe that for p ≫ n cases the recovery of predictors presents more problems. In particular, we can see that all methods are impacted by masking and swamping effects as regards predictor recovery. SPARSE-LTS is the best performer, followed by SNCD-H, RLARS and SNCD-Q. Concerning the outlier detection performance, we can see in Figure 4 that the picture is generally better than in Figure 2, proving that the sample size also has a crucial role for a good performance. SPARSE-LTS represents the best option for large values of m, exhibiting very low error rates. MM is the best option for small values of m: it is doing perfectly with respect to masking, while it is still impacted by swamping effect (around 8%). RLARS is doing well at large m: it features a masking rate around 10% and a swamping rate around 4%. GS is performing like RLARS from the masking side, and quite worse than RLARS from the swamping side. The LTS options have instead a swamping rate close to 0 for large values of m, but a large masking rate compared to others (over 10%) and a swamping rate around 5% for small values of m. The SNCD-H versions of ROBOUT work in general quite better than the SNCD-Q versions for all robust regression methods in this case.

[Figure 3 here]

[Figure 4 here]

Under scenario 2a, the overall picture is very close to the one reported in Figure 1 for scenario 1a. The only difference is that the performance of all methods is slightly better for small values of m. The same considerations can be easily extended to scenario 2b with respect to scenario 1b, both for conditional outlier detection and the recovery of predictors. On the contrary, the results for scenario 2c present some different features compared to scenario 1c. Figure 5 shows that in this case SNCD-H and SNCD-Q work quite badly in recovering the predictors, since they tend to include one more predictor quite often, and the same does RLARS. SPARSE-LTS, instead, works well as m reaches 7. In Figure 6, we can instead see that SPARSE-LTS fails completely the outlier recovery for m = 3,5, and then works almost perfectly. The MM and RLARS options show a masking rate comparable to the SPARSE-LTS as m approaches 13. RLARS presents a swamping rate around 4%, while the MM options stand around 12%. The GS options have instead a masking rate around 15%. In the end, the LTS options fail completely, with a masking rate always over 50% across m.

[Figure 5 here]

[Figure 6 here]

Under scenario 3, which introduces leverage outliers, the results change quite abruptly. Concerning scenario 3a (n > p), the recovery of predictors is again impeccable, but we observe in Figure 7 that all methods are suffering from a severe masking effect for small values of m. At the same time, as m approaches 13, the MM options have a masking rate smaller than 20% and decreasing across m, with a swamping rate around 4%. LTS, the second best, shows instead a systematic gap in terms of masking with respect to MM, but no swamping effect. SPARSE-LTS performance only recovers for large values of m, reaching a masking rate of 20% and a swamping rate of 2%. The GS options perform like the MM ones for small values of m but then finish worse than SPARSE-LTS for large values of m. RLARS is completely inconsistent, with a persistent masking rate over 60%.

[Figure 7 here]

Concerning scenario 3b, we observe that almost all methods work perfectly. No method presents a mean number of masked or swamped predictors over 0.03 (out of 3), apart RLARS for $m = 3$. With respect to conditional outlier detection, we should stress that all methods exhibit masking rates close to zero, apart from GS for small values of m (no more than 7%). Concerning swamping rates, MM is the worst performer with 4%, RLARS stands around 3%, SPARSE-LTS around 2%, GS around 1% and LTS close to 0%.

Concerning scenario 3c, the recovery of predictors works worse, as expected. In particular, RLARS presents a mean number of masked predictors and swamped predictors around 0.1 out of 3. SPARSE-LTS masks around 0.3 out of 3 predictors and swamps less than 0.05 in mean. SNCD-H reaches 0.4 masked predictors and 0.56 swamped predictors on average at $m = 17$. The same figures for SNCD-Q are 0.59 and 0.69 respectively. As regards conditional outlier detection, we observe that all methods are challenged under this scenario. The best performer is SNCD-H+MM, which attains a masking rate slightly above 30% and a swamping rate slightly above 20%. SNCD-Q+MM performs only slightly worse than SNCD-H+MM. SPARSE-LTS performance only recovers a bit for very large values of m. SNCD-H+LTS and SNCD-H+GS show a masking rate around 50% at their best, but the latter method also shows a swamping rate around 10%. Their SNCD-Q counterparts present similar but slightly worse figures. RLARS is still completely out of the game, as its masking rate never goes below 70%.

[Figure 8 here]

[Figure 9 here]

Finally, under the most challenging scenario 4 which combines leverage outliers with sparsity, the recovery of predictors works in a very similar way compared to scenario 3. The only difference is the slightly worse performance of RLARS for scenario 4b (compared to scenario 3b), and the slightly worse masking rate of true predictors for SPARSE-LTS under scenario 4c compared to scenario 3c. The conditional outlier detection performances also resemble the relative standings of scenario 3 in the three subcases, apart from the slightly worse performance of SPARSE-LTS under scenario 4c compared to scenario 3c.

In Table 1, the top performer for each described scenario at m = 19 is highlighted. As it can be seen for almost all cases, variants of our proposed ROBOUT approach represent the best performing methods. For scenario 1, we observe that for large n and small p LTS is the top performer, for intermediate values of p and n MM is the top performer, while for large p and small n SPARSE-LTS recovers as the best, because it retrieves the true predictors better than SNCD. For large p, SNCD-H works better than SNCD-Q for the recovery of predictors. RLARS is instead completely out of the game.

[Table 1 here]

Under scenario 2, we observe the same pattern of scenario 1 for large n and small p, with LTS prevailing over SPARSE-LTS and MM. For intermediate values of p and n (with p > n), MM is again the top performer, followed by LTS and GS, while SPARSE-LTS and RLARS fail completely. For large p, SPARSE-LTS is working properly, followed by RLARS, GS, and MM (which presents a relevant swamping effect). SPARSE-LTS needs a relevant perturbation level to recover outliers, even if it recovers the predictors better than RLARS and SNCD (in the order). For large p, the SNCD-H versions of ROBOUT perform quite better in recovering both predictors and conditional outliers than the respective SNCD-Q counterparts.

Under scenario 3, the situation is a bit different. We see that for large n MM is the top performer, as the leverage outliers impact too much on the other methods. For intermediate p and n, instead, LTS and SPARSE-LTS recover as the best performers, with RLARS performing acceptably.

When p is large, however, the SNCD-H versions of LTS, GS and MM are the best (in the order), followed by the SNCD-Q versions of the same methods, while SPARSE-LTS is not useful unless the perturbation is very strong and RLARS fails completely, in spite of being the best in recovering predictors.

Under scenario 4, the relative performance is similar to the one of scenario 3 for cases a and b, the only exception being RLARS which performs relatively better for case a and SPARSE-LTS for case b. For $p = 500$, $n = 50$, moreover, the hierarchy is strictly SNCD-H+MM, SNCD-Q+MM, SNCD-H+GS, SNCD-Q+GS, SNCD-H+LTS, SNCD-Q+LTS, SPARSE-LTS, RLARS. So, in this case MM and SNCD-H are clearly the best performers in recovering outliers and predictors respectively, even if the performance level is generally quite low. The fact that RLARS is the best performer for predictor recovery, followed by SPARSE-LTS, SNCD-H and SNCD-Q, does not produce any effect on the identification of conditional outliers.

All in all, SNCD-H+MM appears to be the only method which performs acceptably under any scenario. In particular, whenever the masking level of the competitors is not close to 0, the MM options appear to be the best for the limited masking level, in spite of a small persistent swamping effect. In addition, SNCD-H versions are performing better than the respective SNCD-Q counterparts, because SNCD-H is less prone to swamping in the predictor recovery. The relative standings of Scenario 2 are particularly instructive: as p increases, the LTS options are overwhelmed by the masking effect, while SPARSE-LTS presents a relevant masking effect for intermediate values of p and n.

In Tables 2 and 3, we have also reported the means for both the masking and the swamping rate for all methods and across all scenarios and settings at m = 19. We stress that the standard deviations generally follow linearly the mean performance levels. In Table 4, we have instead reported the same statistics for the F1-score, which shows that the MM options of ROBOUT emerge as the most versatile under the different perturbation conditions.

[Table 2 here]

[Table 3 here]

[Table 4 here]

To sum up, in the absence of leverage outliers within the predictors set, SNCD-H+MM appears to be the only safe method for all combinations of p and n, in spite of the slight persistent swamping effect. To reinforce this conclusion, we have also tested a scenario with a higher level of sparsity, equal to 70%, and also another one with a 30% additional level of sparsity into the matrix of predictors. In both cases, the results are absolutely similar to the results of scenario 2. It is important to stress that this superiority of the MM options is particularly relevant for all scenarios with large and intermediate values of both p and n (p > n case).

On top of that, computational times also play a relevant role. In Table 5, we report the mean computational times across ten runs of the eight methods for a typical replicate of scenarios 1a, 1b, 1c. We can see that the MM options are always the fastest, followed by the LTS ones which are still very fast. SPARSE-LTS particularly suffers from a large n, as the running time is about 1000 times larger in that case (scenario 1a). As p increases (scenario 1b), we observe that the GS options and SPARSE-LTS recover a bit, even if they are still more than 100 times slower than LTS and MM, while RLARS is still very slow. The situation of RLARS improves for scenario 1c, even if it is still 100 times slower than MM.

In Table 5, we can also observe that the situation for scenario 2 is pretty similar, apart from scenario 2b, where SPARSE-LTS is particularly slow, and RLARS is relatively faster. A similar pattern is observed for scenarios 3 and 4, as the relative performance among the different methods does not change.

[Table 5 here]

**A real banking supervisory data example**

In this section, we apply the proposed conditional outlier detection procedure, ROBOUT, to a real banking dataset. In particular, we analyze a set of the so-called FINREP templates ('FINREP' stands for 'Financial Reporting') which contain all basic breakdowns of the banks' balance sheet, therefore providing very granular information on banks' activities. These supervisory data are submitted by the European banks to the European Central Bank. Our sample consists of a cross section of the largest 365 banks of the euro area, on which 453 balance sheet indicators are systematically measured. The reference date of the data is end-2014.

The dataset in question is compositional, i.e. it includes some parent categories, like 'Debt', and their sub-parent categories like 'Debt versus other banks', 'Debt versus central government', etc. In addition, many structural zeros, i.e. values which could not be reported by the bank in question, are present. Furthermore, there may be scale effects in the composition of activities, e.g. large banks with more complex strategies may, for example, make more extended use of derivative instruments. Therefore, we decide to normalize each indicator by the total asset of the relative bank. The bulk of the resulting distributions (usually strongly skewed, as pointed out by [49]) is expected to be located within a subset of [0,1], given the normalization with respect to assets and the fact that most items are lower than the bank's size – although a few variables, such as the notional amounts of derivatives, may exceed unity i.e. their size may be higher than the bank's total assets.

Before running our ROBOUT procedure on the described dataset, we provide a statistical picture of its content. We observe that our data matrix presents a sample size n = 365, a dimension p = 453, 69.9% of zero entries, a mean absolute correlation of 0.0424 and a mean correlation of 0.0080. We then set the log of bank's size as the dependent variable. The distribution of the log of banks' size shows that the log-normality assumption on total assets cannot be rejected. The p-value of the Jarque-Bera test is in fact 40.18%. The fact that the size follows a log-normal distribution is intuitive, given the existence of a large number of small- and medium-sized banks, high variance (also reflecting the size dispersion of the countries in the sample e.g. with respect to their GDP) and the non-negativity of the size variable. The mean and the median of the log-size are almost equal (19.71 VS 19.81), while the standard deviation is slightly larger than the MAD (2.14 VS 1.73).

We run the six options of the ROBOUT procedure with α=0.1 and $K = 2,3,4,5$. The results confirm the previously outlined findings. In fact, we observe that:

- the GS options are not feasible for any value of K, due to the excessive number of zero entries.

- SPARSE-LTS and RLARS are not feasible as well, due to their explosive computational time.

- the LTS and the MM options can be computed. In particular, the SNCD-H and SNCD-Q options followed by each robust regression method return the same results for $K = 4,5$.

Following the error rates patterns of Scenario 2 when n~p, we know that LTS is in general more masking oriented, while MM is only slightly swamping oriented. Therefore, we trust the MM options, also because the LTS options return a very small number of conditional outliers in the log-assets.

Looking at the regression models estimated via the MM options, we can judge that $K = 5$ is the most appropriate choice, since the estimated coefficients are all strongly significant (see Table 6), and their signs are meaningful. In particular, we can see that derivatives (both for trading and hedging purposes) at notional amount and debt securities at amortised cost have a positive impact on the log-assets, while hedging derivatives at carrying amount and loans and advances on demand and short notice to credit institutions have a negative impact. The adjusted R-squared overcomes 82%.

[Table 6 here]

The resulting 365 ×5 matrix of predictors has a 29.70% of zero elements, and 1.92% of zero rows. The mean absolute correlation among predictors is 0.3089, the mean signed correlation is 0.1606. In Figure 10, we report the estimated residuals across our sample. We can observe that all the residuals of the 15 conditional outliers are negative, showing that outlying banks in this sense present a too small value of total assets compared to the values they present for the predictors. At the same time, we can observe in Figure 11 that the 15 conditional outliers present an average value for log-assets (apart from the bank reporting the lowest asset). We are therefore able to identify a very different outlyingness dimension with respect to the usual univariate one.

[Figure 10 here]

[Figure 11 here]

[Table 7 here]

The nature of recovered conditional outliers can be further understood by looking at their relative standing in the distribution of recovered predictors. In Table 7, we can understand that the recovered conditional outliers present outstanding values of derivatives (both for trading and hedging) at notional amount. Therefore, the set of recovered outliers mainly identifies small banks whose small size is an outlier when compared to the relatively large amount of trading derivatives present in their balance sheet, thus uncovering an unexpected outlyingness dimension for euro area banks.

**Conclusions**

In this paper, we propose a new conditional outlier detection methodology, called ROBOUT. ROBOUT is very versatile and flexible, as it is able to robustly spot conditional outliers under many different perturbed conditions and different combinations of sample size and dimension. In addition, ROBOUT works efficiently on datasets with many observations and variables as it is computationally lighter than alternative methods, and it is able to robustly select the most relevant predictors to predict any single variable.

ROBOUT presents two options to select relevant predictors, based on the Huber or the quantile loss of residuals, and three options to estimate a robust regression, namely, the well-known LTS, GS and MM methods. The perturbation conditions we have tested include conditional outliers in variance in the response variable, leverage outliers in the predictors, and the presence of many zeros in the data matrix. We have tested the performance of ROBOUT when the dimension is smaller ($p<n$), larger ($p>n$) or much larger ($p \gg n$) than the sample size.

A comprehensive simulation study has shown that the Huber loss works slightly better than the quantile loss to spot predictors, and there is a clear performance hierarchy for what concerns conditional outlier detection: MM is on average the best method, followed by LTS and GS. The existing competitors, namely the RLARS and SPARSE-LTS integrated procedures, usually work quite worse than MM and LTS, particularly for the case p>n. In addition, their computational time is hundred times larger than the ROBOUT options with the MM regression, which are the fastest.

Among the research developments possibly coming from this work, the most interesting one is probably the derivation of an effective integrated procedure based on the Huber loss to spot both predictors and conditional outliers simultaneously. A very recent attempt to derive a similar procedure using LADLASSO was made in [51], where a method called PWLAD-LASSO is successfully implemented to spot regression outliers but only under the case $p < n$, a constraint which is not binding for the method proposed in this paper.

**Declaration of interest**

We have no conflict of interest to disclosure of any kind.

**Data availability statement**

The dataset used in Section 4 is private and confidential, as subject to institutional constraints. For these reasons, it may not made publicly available. However, the paper is completed by two R functions, 'cond_out_gen' and 'robout', which can regenerate all the simulated settings of this paper and compute all the possible variants of ROBOUT procedure.

**References**

[1] Hawkins DM. Identification of outliers. London: Chapman and Hall; 1980 May.

[2] Barnett V, Lewis T. Outliers in statistical data. 3rd edition: Wiley; 1994 May.

[3] Rousseeuw PJ, Hubert M. Anomaly detection by robust statistics. Wiley Interdisciplinary Reviews: Data Mining and Knowledge Discovery. 2018 March; 8(2):e1236.

[4] Alfons A, Croux C, Gelper S. Sparse least trimmed squares regression for analyzing high-dimensional large data sets. The Annals of Applied Statistics. 2013;7(1):226-48.

[5] Wang H. Forward regression for ultra-high dimensional variable screening. Journal of the American Statistical Association. 2009 Dec 1;104(488):1512-24.

[6] Georghiades AS, Belhumeur PN, Kriegman DJ. From few to many: Illumination cone models for face recognition under variable lighting and pose. IEEE transactions on pattern analysis and machine intelligence. 2001 Jun;23(6):643-60.

[7] Rousseeuw PJ, Leroy AM. Robust Regression and Outlier Detection. New York: Wiley-Interscience; 1987.

[8] Aggarwal CC. Outlier analysis. In Data mining 2015 (pp. 237-263). Springer, Cham.

[9] Beyer K, Goldstein J, Ramakrishnan R, Shaft U. When is 'nearest neighbor' meaningful? In International conference on database theory 1999 Jan 10 (pp. 217-235). Springer, Berlin, Heidelberg.

[10] Angiulli F, Basta S, Lodi S, Sartori C. Distributed strategies for mining outliers in large data sets. IEEE Transactions on Knowledge and Data Engineering. 2012 Apr 3;25(7):1520-32.

[11] Angiulli F, Fassetti F. Distance-based outlier queries in data streams: the novel task and algorithms. Data Mining and Knowledge Discovery. 2010 Mar 1;20(2):290-324.

[12] Duan L, Xu L, Liu Y, Lee J. Cluster-based outlier detection. Annals of Operations Research. 2009 Apr 1;168(1):151-68.

[13] Durrant RJ, Kabán A. When is 'nearest neighbour' meaningful: A converse theorem and implications. Journal of Complexity. 2009 Aug 1;25(4):385-97.

[14] Zimek A, Schubert E, Kriegel HP. A survey on unsupervised outlier detection in high-dimensional numerical data. Statistical Analysis and Data Mining: The ASA Data Science Journal. 2012 Oct;5(5):363-87.

[15] Zhou C, Paffenroth RC. Anomaly detection with robust deep autoencoders. In Proceedings of the 23rd ACM SIGKDD International Conference on Knowledge Discovery and Data Mining 2017 Aug 4 (pp. 665-674).

[16] Kriegel HP, Kröger P, Schubert E, Zimek A. LoOP: local outlier probabilities. In Proceedings of the 18th ACM conference on Information and knowledge management 2009 Nov 2 (pp. 1649-1652).

[17] Lazarevic A, Kumar V. Feature bagging for outlier detection. InProceedings of the eleventh ACM SIGKDD international conference on Knowledge discovery in data mining 2005 Aug 21 (pp. 157-166).

[18] Hong C, Hauskrecht M. Multivariate Conditional Anomaly Detection and Its Clinical Application. In AAAI 2015 Jan 25 (pp. 4239-4240).

[19] Pang G, Cao L, Chen L, Lian D, Liu H. Sparse Modeling-Based Sequential Ensemble Learning for Effective Outlier Detection in High-Dimensional Numeric Data. In AAAI 2018 Jan 1 (pp. 3892-3899).

[20] Varian HR. Big data: New tricks for econometrics. Journal of Economic Perspectives. 2014 May;28(2):3-28.

[21] Tibshirani R. Regression shrinkage and selection via the lasso. Journal of the Royal Statistical Society: Series B (Methodological). 1996 Jan;58(1):267-88.

[22] Mandal BN, Ma J. l1 regularized multiplicative iterative path algorithm for non-negative generalized linear models. Computational Statistics & Data Analysis. 2016 Sep 1;101:289-99.

[23] Tibshirani R, Saunders M, Rosset S, Zhu J, Knight K. Sparsity and smoothness via the fused lasso. Journal of the Royal Statistical Society: Series B (Statistical Methodology). 2005 Feb;67(1):91-108.

[24] Meinshausen N, Bühlmann P. High-dimensional graphs and variable selection with the lasso. The annals of statistics. 2006;34(3):1436-62.

[25] Yuan M, Lin Y. Model selection and estimation in regression with grouped variables. Journal of the Royal Statistical Society: Series B (Statistical Methodology). 2006 Feb;68(1):49-67.

[26] Simon N, Friedman J, Hastie T, Tibshirani R. A sparse-group lasso. Journal of computational and graphical statistics. 2013 Apr 1;22(2):231-45.

[27] Zou H. The adaptive lasso and its oracle properties. Journal of the American statistical association. 2006 Dec 1;101(476):1418-29.

[28] Fan J, Lv J. Sure independence screening for ultrahigh dimensional feature space. Journal of the Royal Statistical Society: Series B (Statistical Methodology). 2008 Nov 1;70(5):849-911.

[29] Kim Y, Choi H, Oh HS. Smoothly clipped absolute deviation on high dimensions. Journal of the American Statistical Association. 2008 Dec 1;103(484):1665-73.

[30] Candes E, Tao T. The Dantzig selector: Statistical estimation when p is much larger than n. The Annals of Statistics. 2007;35(6):2313-51.

[31] Bogdan M, Van Den Berg E, Sabatti C, Su W, Candès EJ. SLOPE—adaptive variable selection via convex optimization. The Annals of applied statistics. 2015;9(3):1103.

[32] Lozano AC, Meinshausen N, Yang E. Minimum distance lasso for robust high-dimensional regression. Electronic Journal of Statistics. 2016;10(1):1296-340.

[33] Wang H, Li G, Jiang G. Robust regression shrinkage and consistent variable selection through the LAD-Lasso. Journal of Business & Economic Statistics. 2007 Jul 1;25(3):347-55.

[34] Park H. Outlier-resistant high-dimensional regression modelling based on distribution-free outlier detection and tuning parameter selection. Journal of Statistical Computation and Simulation. 2017 Jun 13;87(9):1799-812.

[35] Wang T, Li Q, Chen B, Li Z. Multiple outliers detection in sparse high-dimensional regression. Journal of Statistical Computation and Simulation. 2018 Jan 2;88(1):89-107.

[36] Rousseeuw PJ. Least median of squares regression. Journal of the American statistical association. 1984 Dec 1;79(388):871-80.

[37] She Y, Owen AB. Outlier detection using nonconvex penalized regression. Journal of the American Statistical Association. 2011 Jun 1;106(494):626-39.

[38] Hawkins DM, Olive DJ. Inconsistency of resampling algorithms for high-breakdown regression estimators and a new algorithm. Journal of the American Statistical Association. 2002 Mar 1;97(457):136-59.

[39] Rousseeuw PJ, Driessen KV. A fast algorithm for the minimum covariance determinant estimator. Technometrics. 1999 Aug 1;41(3):212-23.

[40] Croux C, Rousseeuw PJ, Hössjer O. Generalized S-estimators. Journal of the American Statistical Association. 1994 Dec 1;89(428):1271-81.

[41] Yohai VJ. High breakdown-point and high efficiency robust estimates for regression. The Annals of Statistics. 1987 Jun 1:642-56.

[42] Jurczyk T. Outlier detection under multicollinearity. Journal of Statistical Computation and Simulation. 2012 Feb 1;82(2):261-78.

[43] Khan JA, Van Aelst S, Zamar RH. Robust linear model selection based on least angle regression. Journal of the American Statistical Association. 2007 Dec 1;102(480):1289-99.

[44] Efron B, Hastie T, Johnstone I, Tibshirani R. Least angle regression. The Annals of statistics. 2004;32(2):407-99.

[45] Yi C, Huang J. Semismooth newton coordinate descent algorithm for elastic-net penalized huber loss regression and quantile regression. Journal of Computational and Graphical Statistics. 2017 Jul 3;26(3):547-57.

[46] Rousseeuw PJ, Van Driessen K. Computing LTS regression for large data sets. Data mining and knowledge discovery. 2006 Jan 1;12(1):29-45.

[47] Salibian-Barrera M, Yohai VJ. A fast algorithm for S-regression estimates. Journal of computational and Graphical Statistics. 2006 Jun 1;15(2):414-27.

[48] Koller M, Stahel WA. Sharpening wald-type inference in robust regression for small samples. Computational Statistics & Data Analysis. 2011 Aug 1;55(8):2504-15.

[49] Pison G, Van Aelst S, Willems G. Small sample corrections for LTS and MCD. Metrika. 2002 Apr 1;55(1-2):111-23.

[50] Hubert M, Van der Veeken S. Outlier detection for skewed data. Journal of Chemometrics: A Journal of the Chemometrics Society. 2008 Mar;22(3-4):235-46.

[51] Jiang Y, Wang Y, Zhang J, Xie B, Liao J, Liao W. Outlier detection and robust variable selection via the penalized weighted LAD-LASSO method. Journal of Applied Statistics. 2020 Feb 4:1-3.

**Appendix**

*Further perturbed scenarios*

In this section, we present some more cases of our simulation study, in order to offer a broader view of ROBOUT performance. First of all, we now define the regression model as follows

$$y_i = a_i + \mathbf{X}_{D,i}\boldsymbol{\beta} + \varepsilon_i. \tag{A.1}$$

For $i^* \in O$, we generate conditional outliers in mean by imposing $a_i = 10m$, with $m > 1$, while for $i^* \notin O$, we set $a_i = 10$. In general, we then set $\mathbf{X}_{D,i^*}' \sim N(\mathbf{0}, \mathbf{I})$ for $i^* \in \{1, \cdots, n\}$. If leverage outliers are also allowed, we set instead $\mathbf{X}_{D,i^*}' \sim N(\mathbf{0}, m\mathbf{I})$ for $i^* \in O$.

Another new model setting that we test here allows for multicollinearity in the data matrix combined with sparsity. In particular, we allow for multicollinearity by setting $COV(X_{j*}, X_{j**})=\rho$, $\rho \in [0,1[$, $\forall j^*, j^{**} \in \{1, \cdots, p\}, j^* \neq j^{**}$ in model (2) of Section 2. We again assume that $X_D$, i.e. the data matrix without predictors, is element-wise sparse with $\gamma \times n \times (p - K)$ randomly positioned non-zeros (with $\gamma \in [0,1)$).

*Simulation results*

In this Section, we consider scenario 5, whose data generating process is model (2) in Section 2 with a multicollinearity level $\rho=0.7$ and a sparsity parameter $\gamma=0.3$.

Like in Section 3, we produce three different combinations of p and n:

    a. $p = 100$, $n = 200$;

    b. $p = 200$, $n = 100$;

    c. $p = 500$, $n = 50$.

What is more, we also test model (A.1) without (scenario 6b) and with (scenario 7b) leverage outliers with $p = 100$ and $n = 200$.

In Figure A.1, we observe the performance of the different methods under scenario 6b. We observe that SNCD-Q+LTS, SNCD-Q+MM, and SPARSE-LTS are the best performers in this case, while RLARS, SNCD-H+GS, SNCD-Q+GS present relevant masking problems when m is small. At the same time, SNCD-Q, SNCD-H, SPARSE-LTS and RLARS present no problems in the recovery of predictors.

[Figure A.1 here]

[Figure A.2 here]

In Figure A.2, we report the outlier detection measures for scenario 7b. For small m, we observe that SNCD-H+GS, SNCD-Q+GS and SNCD-H+MM present some slight masking problems for outlier detection, while SPARSE-LTS, SNCD-H+LTS, and SNCD-Q+LTS are excellent. The mean number of masked and swamped predictors stands below 0.05 in all cases (apart from RLARS with m = 3).

Scenario 5, with sparsity in the data matrix and collinearity among variables, shows that the p < n and the p > n cases present exactly the same results. RLARS is not feasible in this case, and the predictors are perfectly recovered. Concerning conditional outlier detection, SNCD-H+LTS is the best performer, followed by SNCD-H+MM and SNCD-Q+MM. SNCD-H+GS and SNCD-Q+GS work quite worse, while SPARSE-LTS fails completely until m is very large.

Under scenario 5c, instead, all the performance measures about the recovery of predictors and conditional outlier detection measures are irremediably corrupted. SPARSE-LTS shows a mean number of masked predictors around 1 out of 3, while SNCD-H and SNCD-Q show an increasing shape for both measures which attains 0.9 (see Figure A.3). The same methods also overcome a mean number of swamped predictors equal to 1 and 1.2 respectively, while SPARSE-LTS stands around 0.2. Concerning conditional outlier detection (Figure A.4), the MM options show a masking rate around 50%, the LTS options and SPARSE-LTS around 70%. The best option in this case is represented by the GS options, whose masking rate decreases below 30% at large m. The swamping rate of the MM options and SPARSE-LTS stands around 30%.

[Figure A.3 here]

[Figure A.4 here]

In Table A.1, we can see that the computational times for scenarios 6b and 7b are smaller for the LTS options, followed by the MM ones. RLARS and GS are still more than 100 times slower than MM, SPARSE-LTS more than 1000 times slower.

For scenario 5, we observe in the same Table that the MM options are around three times faster than the LTS ones, SNCD-Q is faster than SNCD-H for LTS, GS performs from 600 to 40 times worse than MM as p increases, and SPARSE-LTS has a slowness peak for the b scenario, thus being by far the slowest method, even for large p.

[Table A.1 here]

**Tables with captions**

Table 1. Conditional outlier detection: top performer method across all scenarios and settings.

| | | Leverage | |
|---|---|---|---|
| | **Sparsity** | No | Yes |
| **n=200, p=100** | No | SNCD-H+LTS SNCD-Q+LTS | SNCD-H+MM SNCD-Q+MM |
| | Yes | SNCD-H+LTS SNCD-Q+LTS | SNCD-H+MM SNCD-Q+MM |
| **n=100, p=200** | No | SNCD-H+MM | SNCD-H+LTS SNCD-Q+LTS |
| | Yes | SNCD-H+MM | SNCD-H+LTS SNCD-Q+LTS |
| **n=50, p=500** | No | SPARSE-LTS | SNCD-H+MM |
| | Yes | SPARSE-LTS | SNCD-H+MM |

Table 2. Means of the measured masking rates across all scenarios and settings at m = 19 for each method.

| | 1a | 1b | 1c | 2a | 2b | 2c | 3a | 3b | 3c | 4a | 4b | 4c |
|---|---|---|---|---|---|---|---|---|---|---|---|---|
| SNCD-H+LTS | 0.013 | 0.219 | 0.124 | 0.006 | 0.189 | 0.470 | 0.173 | 0.000 | 0.472 | 0.173 | 0.000 | 0.416 |
| SNCD-H+GS | 0.043 | 0.283 | 0.074 | 0.038 | 0.265 | 0.102 | 0.233 | 0.003 | 0.464 | 0.237 | 0.005 | 0.400 |
| SNCD-H+MM | 0.010 | 0.166 | 0.020 | 0.005 | 0.145 | 0.040 | 0.133 | 0.000 | 0.336 | 0.133 | 0.000 | 0.300 |
| SNCD-Q+LTS | 0.013 | 0.244 | 0.144 | 0.006 | 0.189 | 0.508 | 0.173 | 0.000 | 0.520 | 0.173 | 0.000 | 0.456 |
| SNCD-Q+GS | 0.051 | 0.301 | 0.082 | 0.036 | 0.263 | 0.133 | 0.227 | 0.006 | 0.494 | 0.234 | 0.000 | 0.437 |
| SNCD-Q+MM | 0.010 | 0.179 | 0.030 | 0.005 | 0.145 | 0.044 | 0.133 | 0.000 | 0.372 | 0.133 | 0.000 | 0.330 |
| SPARSE-LTS | 0.016 | 0.298 | 0.042 | 0.008 | 0.261 | 0.020 | 0.200 | 0.000 | 0.406 | 0.200 | 0.000 | 0.434 |
| RLARS | 0.046 | 0.664 | 0.046 | 0.052 | 0.664 | 0.046 | 0.660 | 0.000 | 0.708 | 0.660 | 0.000 | 0.710 |

Table 3. Means of the measured swamping rates across all scenarios and settings at m = 19 for each method.

| | 1a | 1b | 1c | 2a | 2b | 2c | 3a | 3b | 3c | 4a | 4b | 4c |
|---|---|---|---|---|---|---|---|---|---|---|---|---|
| SNCD-H+LTS | 0.0009 | 0.0023 | 0.0033 | 0.0015 | 0.0023 | 0 | 0.0011 | 0.0055 | 0.0078 | 0.0011 | 0.0055 | 0.0095 |
| SNCD-H+GS | 0.0909 | 0.0183 | 0.0493 | 0.118 | 0.0185 | 0.0369 | 0 | 0.0093 | 0.0742 | 0.0278 | 0.0185 | 0.0614 |
| SNCD-H+MM | 0.0359 | 0.0532 | 0.066 | 0.0419 | 0.053 | 0.1198 | 0.0441 | 0.048 | 0.2115 | 0.0441 | 0.048 | 0.1374 |
| SNCD-Q+LTS | 0.0009 | 0.0023 | 0.0017 | 0.0015 | 0.0023 | 0 | 0.0011 | 0.0055 | 0.0078 | 0.0011 | 0.0055 | 0.0095 |
| SNCD-Q+GS | 0.1367 | 0.0091 | 0.0498 | 0.0637 | 0.0371 | 0.0661 | 0.0273 | 0.0276 | 0.0789 | 0.0184 | 0 | 0.0567 |
| SNCD-Q+MM | 0.0359 | 0.0614 | 0.0687 | 0.0419 | 0.053 | 0.1169 | 0.0441 | 0.048 | 0.2288 | 0.0441 | 0.0497 | 0.1414 |
| SPARSE-LTS | 0.0174 | 0.0587 | 0.0215 | 0.0124 | 0.036 | 0.0222 | 0.0234 | 0.0144 | 0.1087 | 0.0234 | 0.0144 | 0.116 |
| RLARS | 0.027 | 0.142 | 0.0282 | 0.0315 | 0.142 | 0.0587 | 0.1425 | 0.0312 | 0.0795 | 0.1425 | 0.0232 | 0.0845 |

Table 4. Means of the measured F1-scores across all scenarios and settings at m = 19 for each method.

| | 1a | 1b | 1c | 2a | 2b | 2c | 3a | 3b | 3c | 4a | 4b | 4c |
|---|---|---|---|---|---|---|---|---|---|---|---|---|
| SNCD-H+LTS | 0.993 | 0.851 | 0.906 | 0.996 | 0.885 | 0.609 | 0.993 | 0.997 | 0.906 | 0.996 | 0.885 | 0.609 |
| SNCD-H+GS | 0.900 | 0.809 | 0.919 | 0.878 | 0.827 | 0.914 | 0.900 | 0.991 | 0.919 | 0.878 | 0.827 | 0.914 |
| SNCD-H+MM | 0.976 | 0.874 | 0.951 | 0.976 | 0.892 | 0.912 | 0.976 | 0.974 | 0.951 | 0.976 | 0.892 | 0.912 |
| SNCD-Q+LTS | 0.993 | 0.824 | 0.890 | 0.996 | 0.885 | 0.575 | 0.993 | 0.997 | 0.890 | 0.996 | 0.885 | 0.575 |
| SNCD-Q+GS | 0.858 | 0.801 | 0.916 | 0.927 | 0.814 | 0.876 | 0.858 | 0.974 | 0.916 | 0.927 | 0.814 | 0.876 |
| SNCD-Q+MM | 0.976 | 0.862 | 0.945 | 0.976 | 0.892 | 0.910 | 0.976 | 0.974 | 0.945 | 0.976 | 0.892 | 0.910 |
| SPARSE-LTS | 0.983 | 0.795 | 0.963 | 0.990 | 0.827 | 0.977 | 0.983 | 0.992 | 0.963 | 0.990 | 0.827 | 0.977 |
| RLARS | 0.963 | 0.460 | 0.958 | 0.957 | 0.460 | 0.943 | 0.963 | 0.984 | 0.958 | 0.957 | 0.460 | 0.943 |

Table 5. Typical computational times in seconds (Intel CORE i7 16 GHz RAM) for all methods and across all scenarios and settings.

|  | Scenario 1a | Scenario 1b | Scenario 1c | Scenario 2a | Scenario 2b | Scenario 2c |
|---|---|---|---|---|---|---|
| **SNCD-H+LTS** | 0.032 | 0.052 | 0.059 | 0.075 | 0.041 | 0.051 |
| **SNCD-Q+LTS** | 0.017 | 0.040 | 0.023 | 0.047 | 0.024 | 0.015 |
| **SNCD-H+GS** | 1.485 | 6.093 | 1.447 | 16.057 | 2.255 | 0.661 |
| **SNCD-Q+GS** | 1.522 | 6.062 | 1.403 | 15.860 | 2.207 | 0.645 |
| **SNCD-H+MM** | 0.010 | 0.019 | 0.012 | 0.020 | 0.015 | 0.010 |
| **SNCD-Q+MM** | 0.010 | 0.019 | 0.012 | 0.020 | 0.014 | 0.010 |
| **SPARSE-LTS** | 23.704 | 4.316 | 1.884 | 12.302 | 36.639 | 1.761 |
| **RLARS** | 2.789 | 45.049 | 0.545 | 41.998 | 4.120 | 0.489 |
|  | Scenario 3a | Scenario 3b | Scenario 3c | Scenario 4a | Scenario 4b | Scenario 4c |
| **SNCD-H+LTS** | 0.056 | 0.071 | 0.029 | 0.056 | 0.033 | 0.020 |
| **SNCD-Q+LTS** | 0.040 | 0.027 | 0.013 | 0.041 | 0.016 | 0.003 |
| **SNCD-H+GS** | 6.299 | 2.990 | 0.729 | 7.373 | 1.690 | 0.162 |
| **SNCD-Q+GS** | 6.678 | 2.954 | 0.753 | 7.558 | 1.757 | 0.082 |
| **SNCD-H+MM** | 0.022 | 0.016 | 0.011 | 0.024 | 0.010 | 0.001 |
| **SNCD-Q+MM** | 0.022 | 0.018 | 0.010 | 0.023 | 0.010 | 0.001 |
| **SPARSE-LTS** | 5.001 | 26.088 | 1.826 | 5.065 | 18.466 | 0.340 |
| **RLARS** | 48.192 | 4.582 | 0.540 | 30.970 | 2.250 | 0.075 |

Table 6. Banking data: estimated MM regression output, with number of regressors K = 5.

| Identifier | Name | Estimate | Std. error | t-value | p-value |
|---|---|---|---|---|---|
| | Intercept | 18.9432 | 0.1564 | 121.115 | <2E-16 |
| {'F0101_r240_c010'} | Derivatives – Hedge accounting – Carrying amount | -43.0174 | 13.9705 | -3.079 | 0.00224 |
| {'F0500_r010_c030'} | Loans and advances on demand [call] and short notice [current account] – Credit institutions | -4.7105 | 0.9455 | -4.982 | 9.81E-07 |
| {'F0801a_r360_c030'} | Debt securities issued – Amortised cost | 4.3541 | 1.2376 | 3.518 | 0.00049 |
| {'F1000_r320_c030'} | Derivatives: Trading – OTC rest – Notional amount – Total trading | 7.8773 | 0.853 | 9.235 | <2E-16 |
| {'F1101_r500_c030'} | Derivatives – Hedge accounting – Notional amount – Total Hedging | 4.2476 | 0.7814 | 5.436 | 1.01E-07 |
| | | | | | |
| **Robust residual standard error:** | | 1.52 | | | |
| **Multiple R-squared:** | | 0.8241 | | | |
| **Adjusted R-squared:** | | 0.8216 | | | |

Table 7. Median and median absolute deviation (MAD) of log-assets and the five selected predictors for outliers and non-outliers.

|  |  | log – size | Derivatives – Hedge accounting – Carrying amount | Loans and advances on demand [call] and short notice [current account] – Credit institutions | Debt securities issued – Amortised cost | Derivatives: Trading – OTC rest – Notional amount – Total trading | Derivatives – Hedge accounting – Notional amount – Total Hedging |
|---|---|---|---|---|---|---|---|
| **Non – outliers** | median | 19.8104 | 0.0002 | 0.0132 | 0.0387 | 0.0078 | 0.0481 |
| **Outliers** | median | 19.9485 | 0.0023 | 0.0135 | 0.1046 | 0.5608 | 0.1498 |
| **Non – outliers** | MAD | 1.7312 | 0.0067 | 0.0562 | 0.0872 | 0.1036 | 0.1601 |
| **Outliers** | MAD | 1.6705 | 0.0044 | 0.0369 | 0.121 | 0.3941 | 0.3388 |

Table A.1. Computational times for all methods across scenarios 5,6b,7b.

|  | scenario 5a | scenario 5b | scenario 5c | scenario 6 | scenario 7 |
|---|---|---|---|---|---|
| **SNCD-H+LTS** | 0.0470 | 0.0309 | 0.0248 | 0.0510 | 0.0390 |
| **SNCD-Q+LTS** | 0.0342 | 0.0193 | 0.0108 | 0.0181 | 0.0196 |
| **SNCD-H+GS** | 6.3277 | 1.6626 | 0.4050 | 2.9042 | 2.0011 |
| **SNCD-Q+GS** | 5.9582 | 1.6832 | 0.4054 | 3.3373 | 1.9074 |
| **SNCD-H+MM** | 0.0169 | 0.0101 | 0.0089 | 0.0124 | 0.0114 |
| **SNCD-Q+MM** | 0.0165 | 0.0170 | 0.0073 | 0.0109 | 0.0117 |
| **SPARSE-LTS** | 5.5845 | 27.4942 | 1.2536 | 26.9204 | 36.3760 |
| **RLARS** |  |  |  | 2.9618 | 2.6658 |

**Figures**

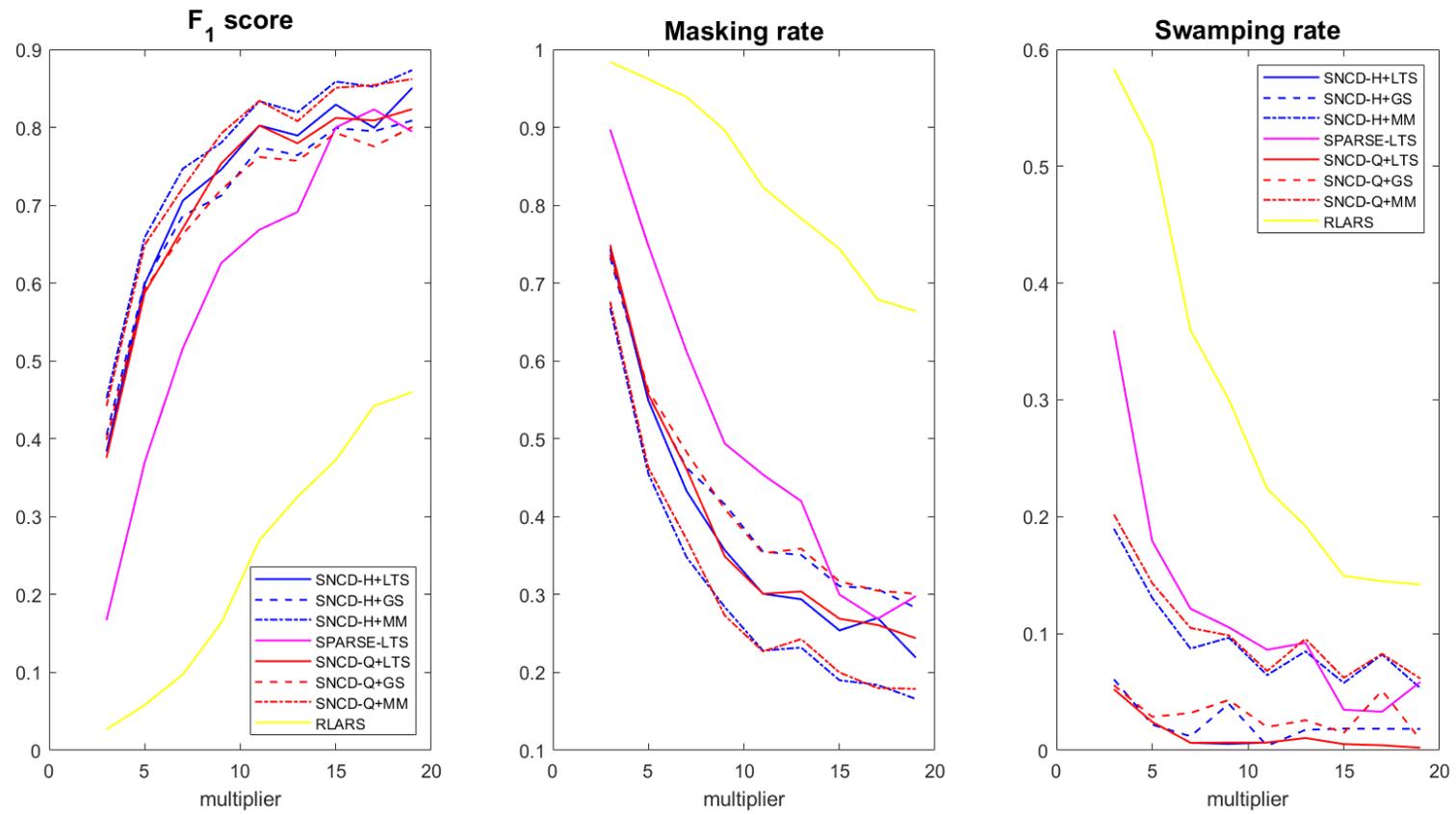

Figure 1.

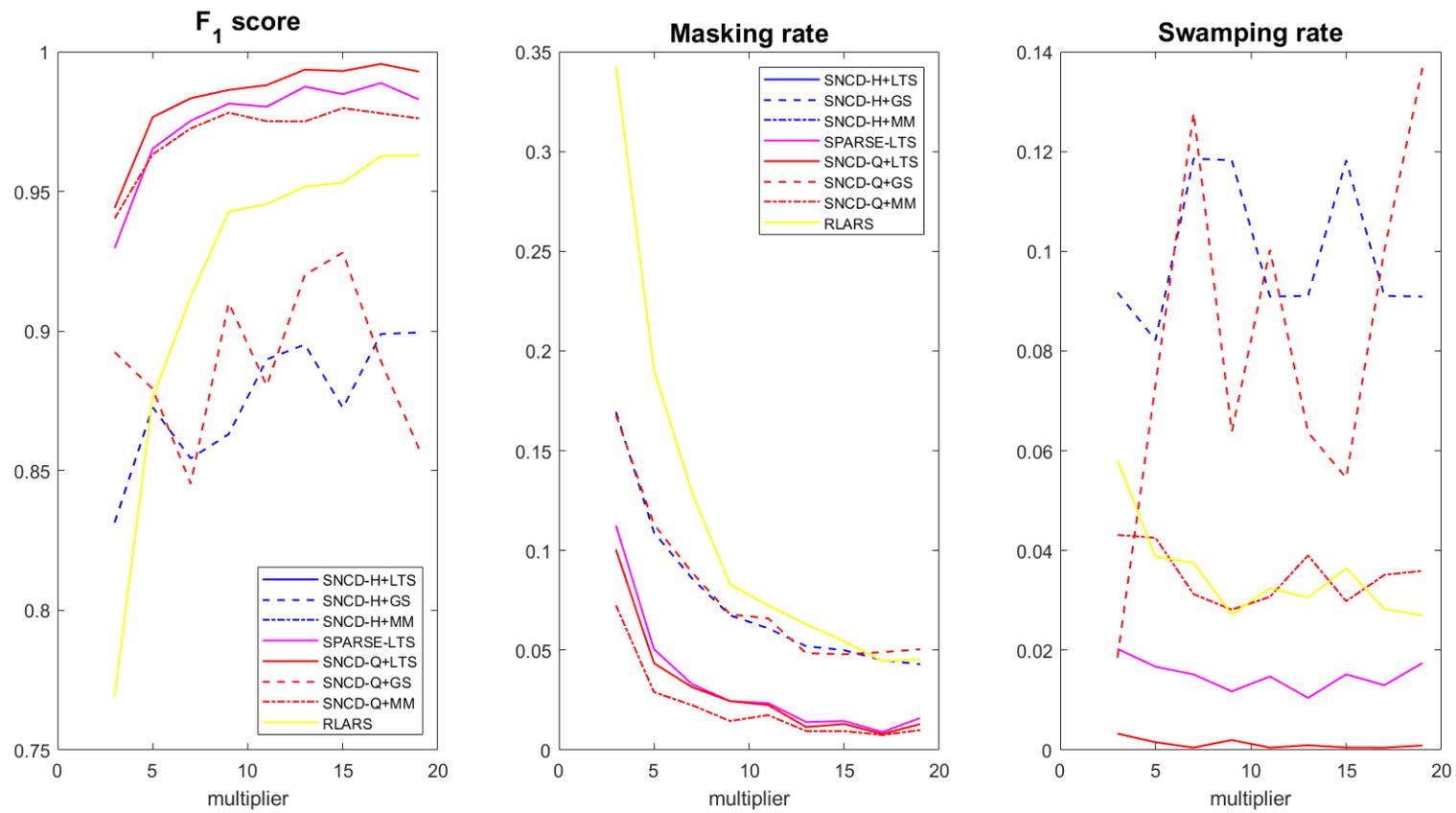

Figure 2.

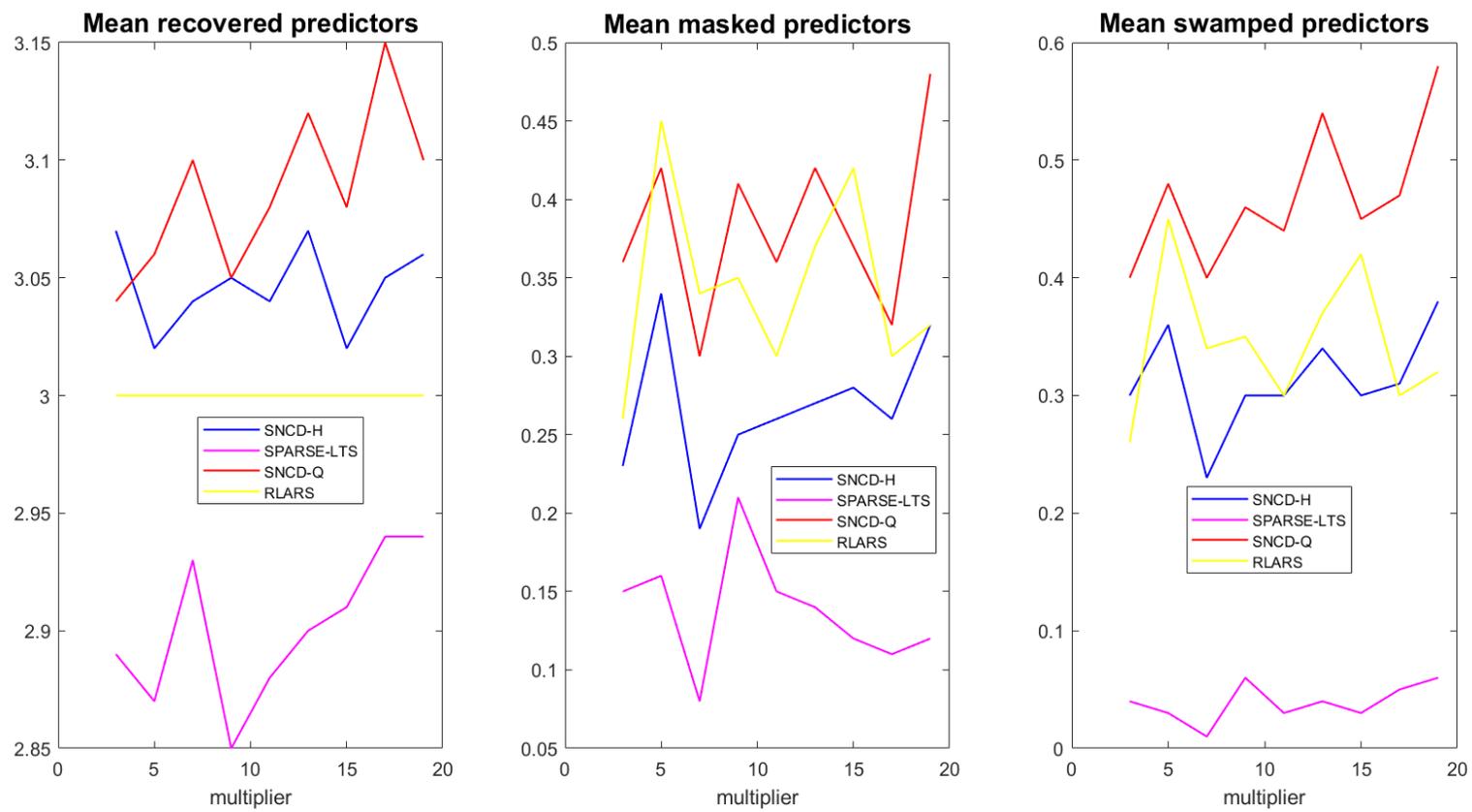

Figure 3.

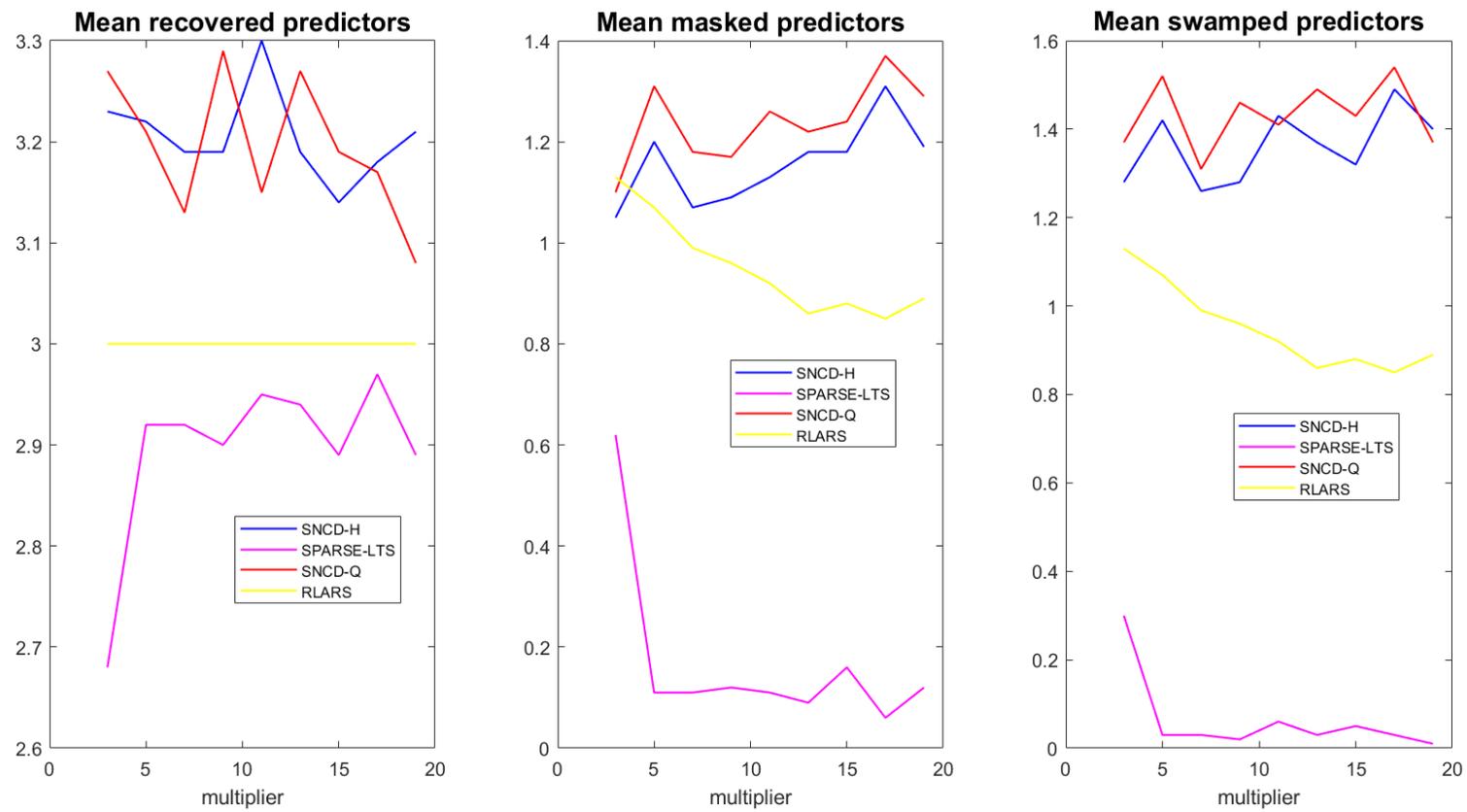

Figure 4.

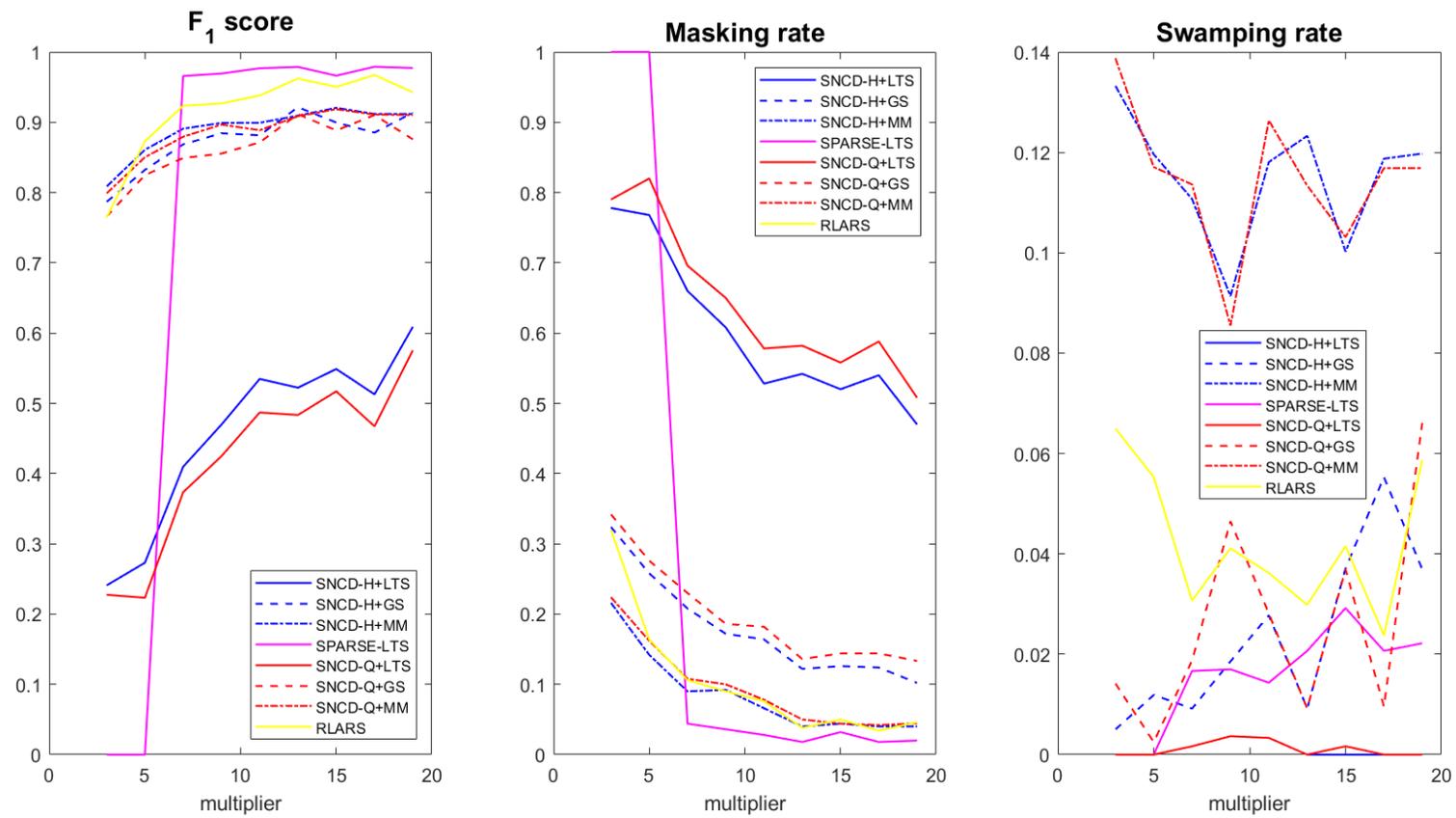

Figure 5.

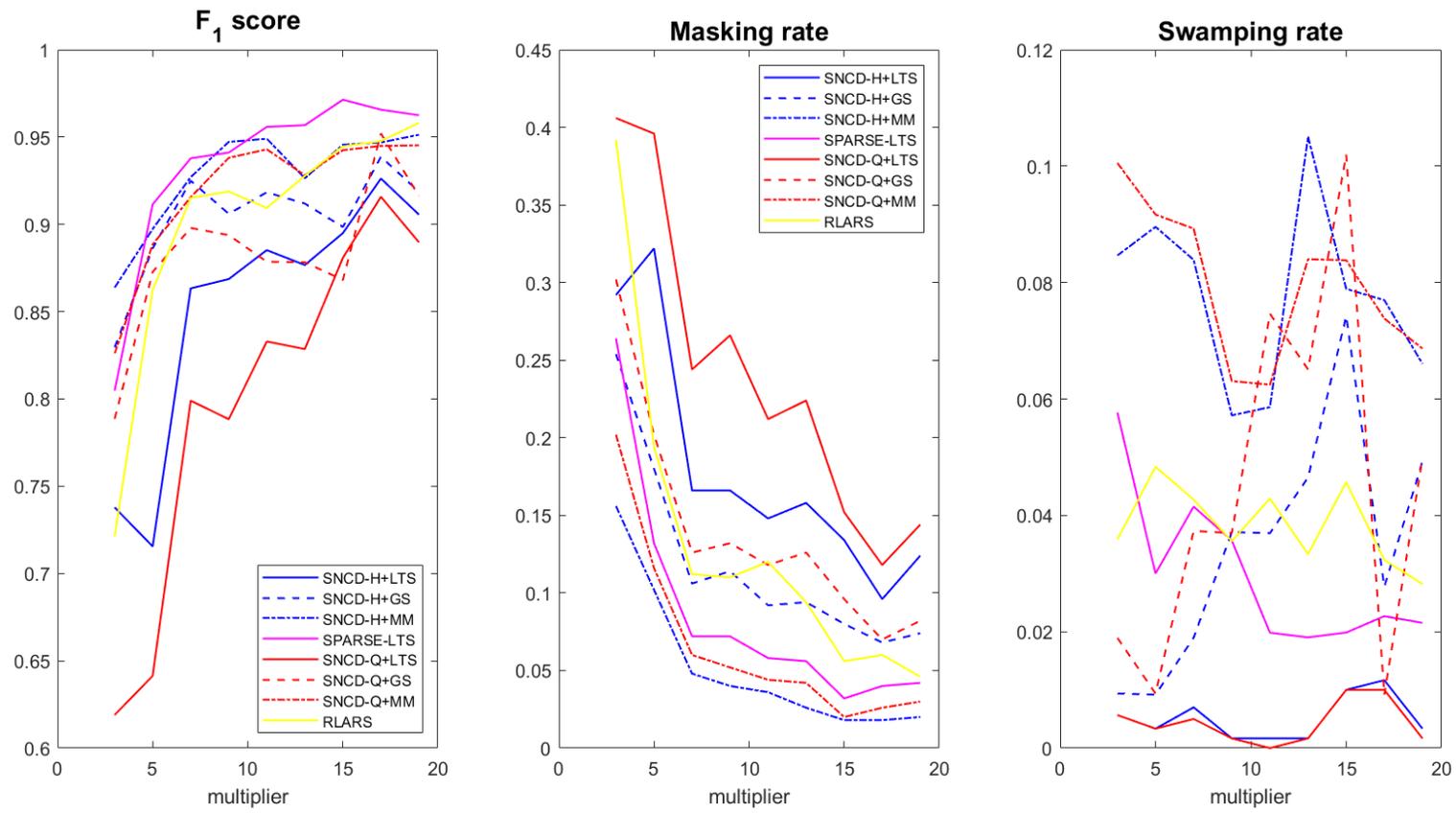

Figure 6.

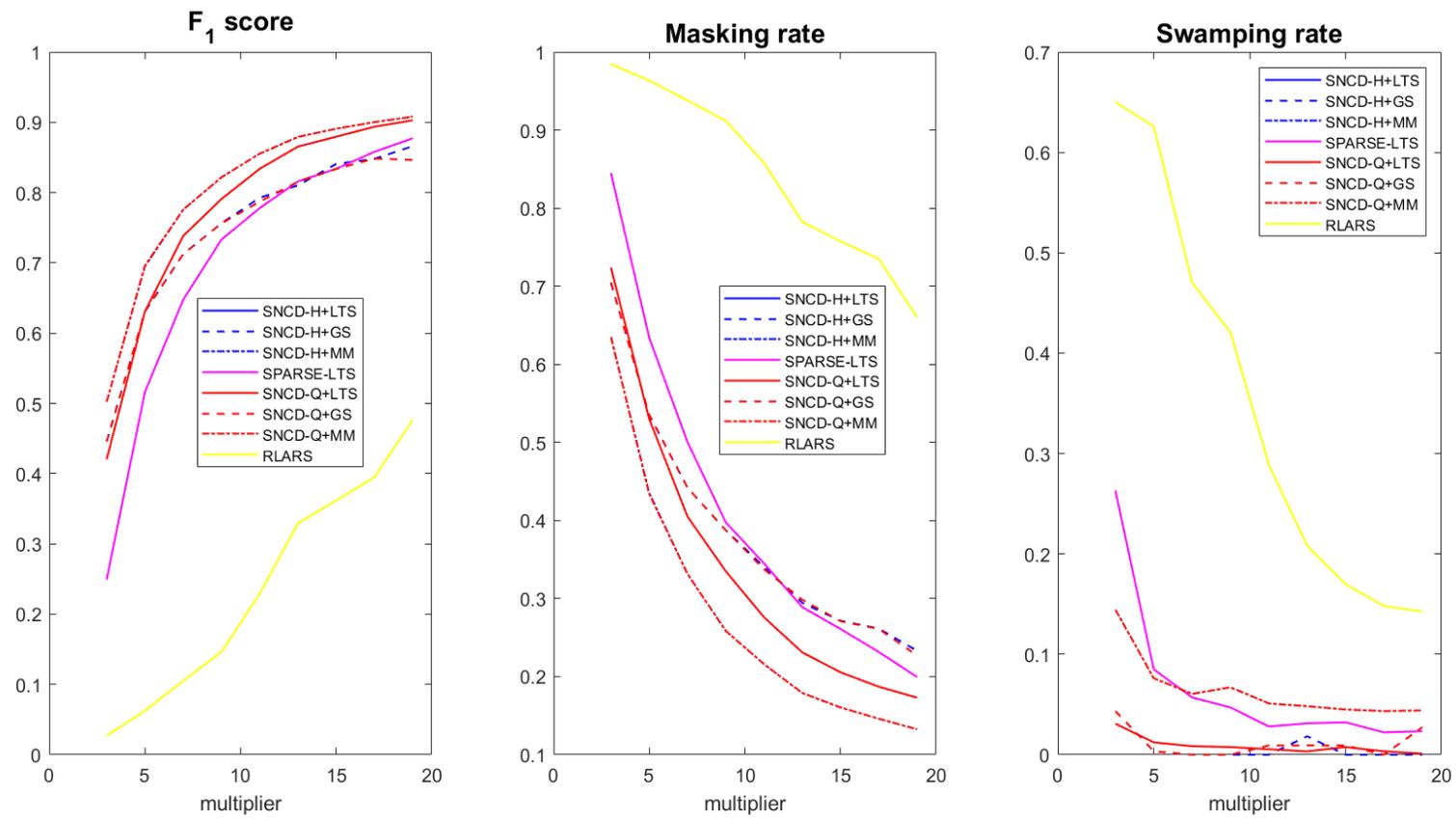

Figure 7.

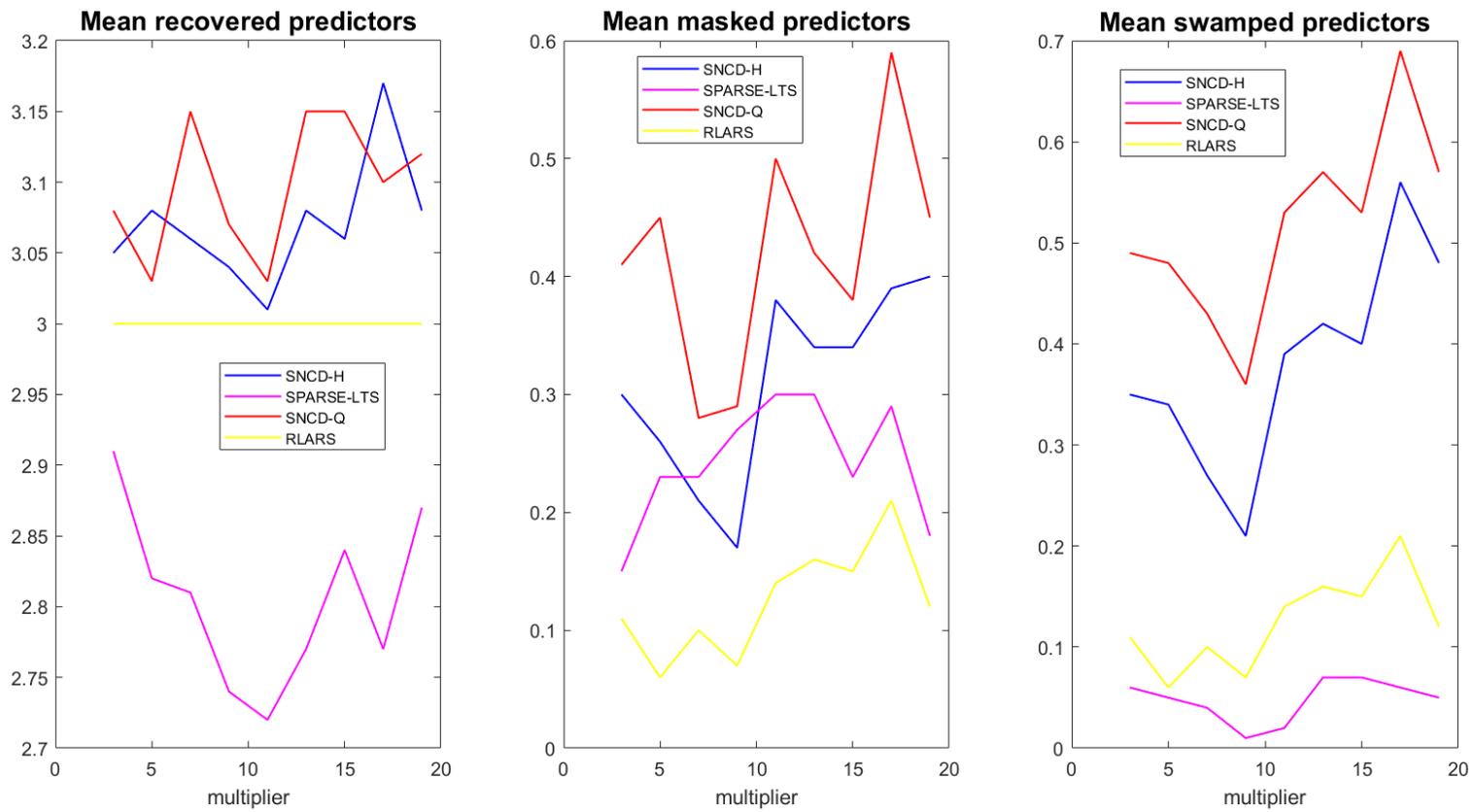

Figure 8.

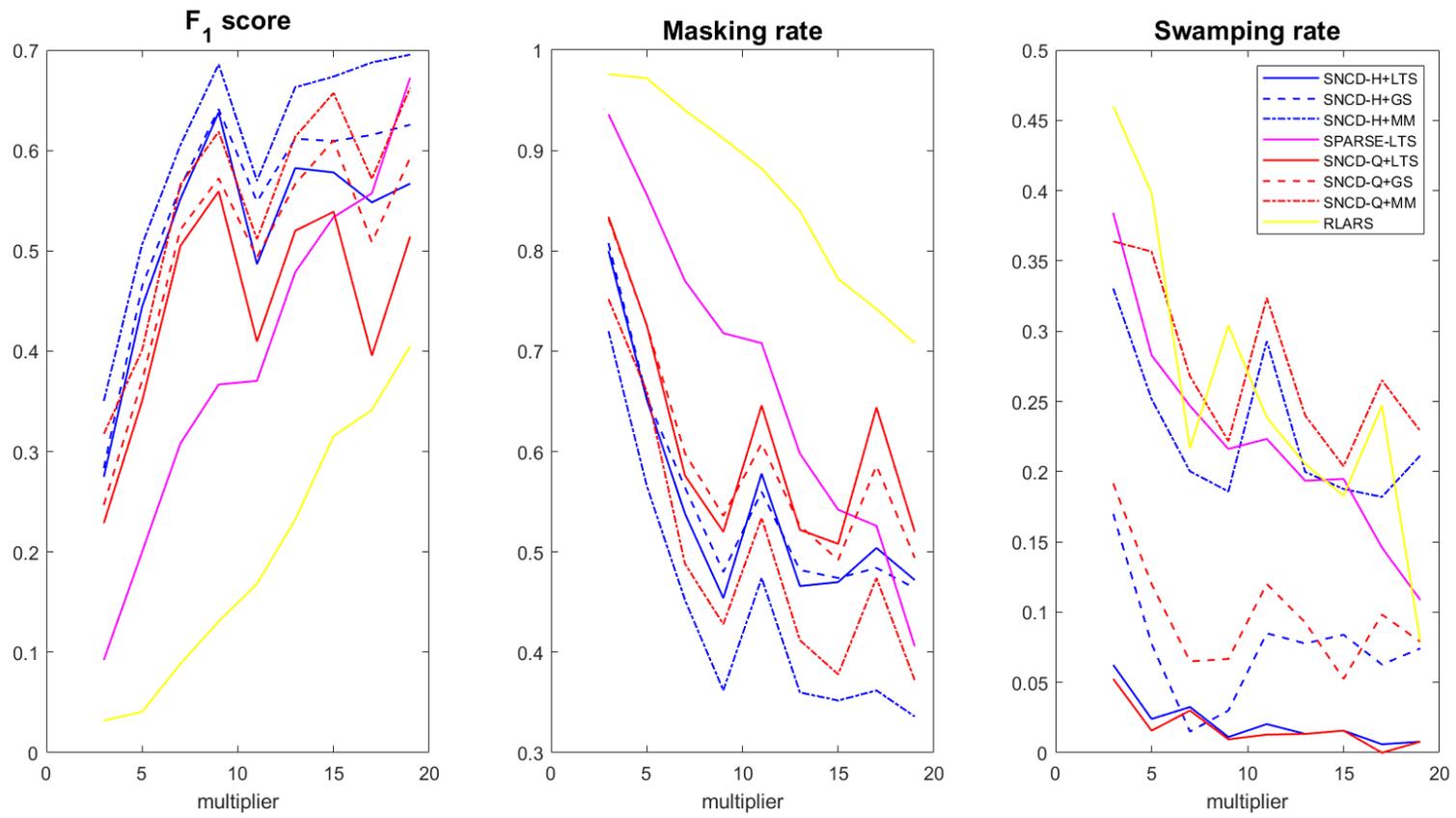

Figure 9.

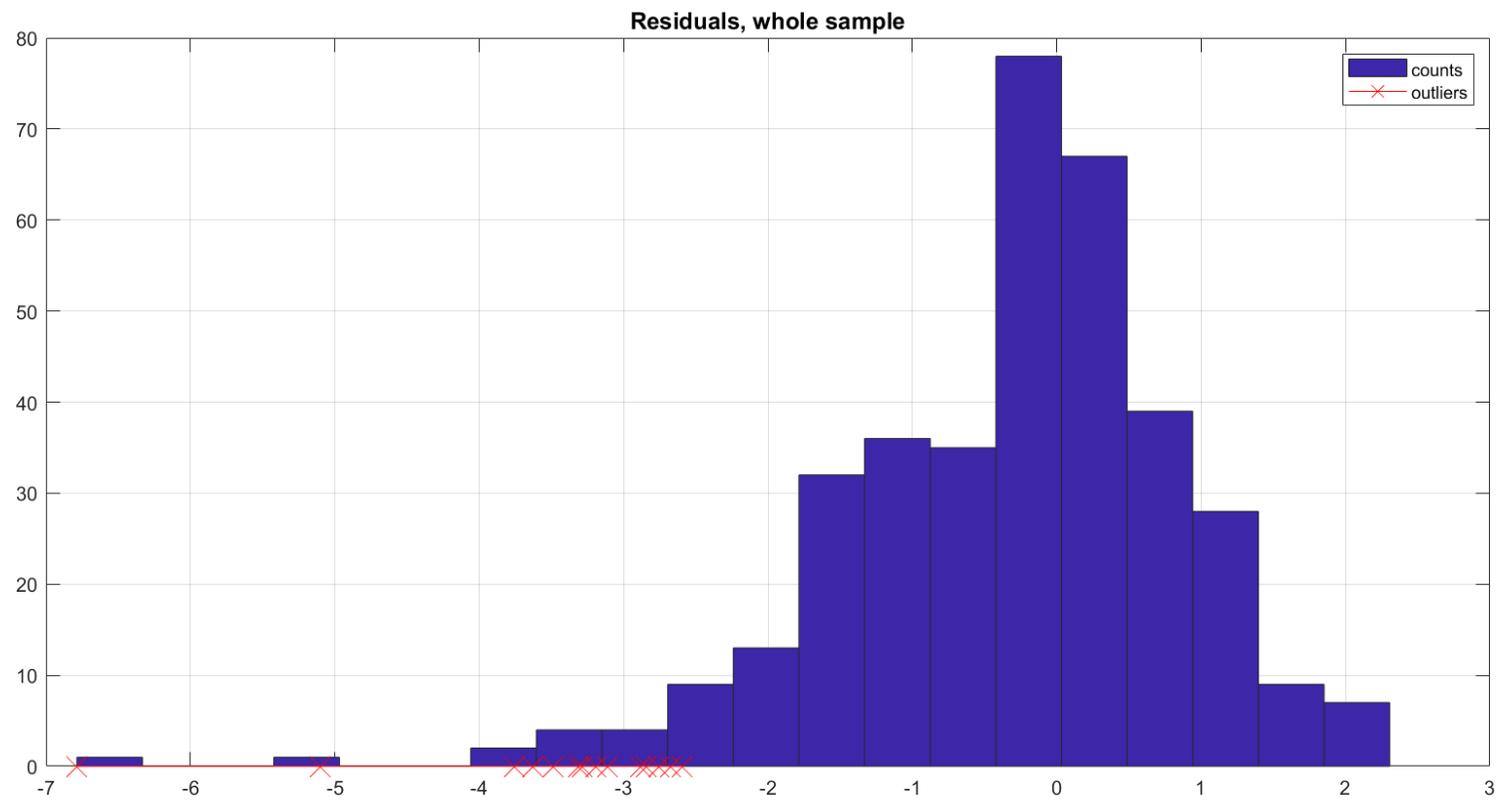

Figure 10.

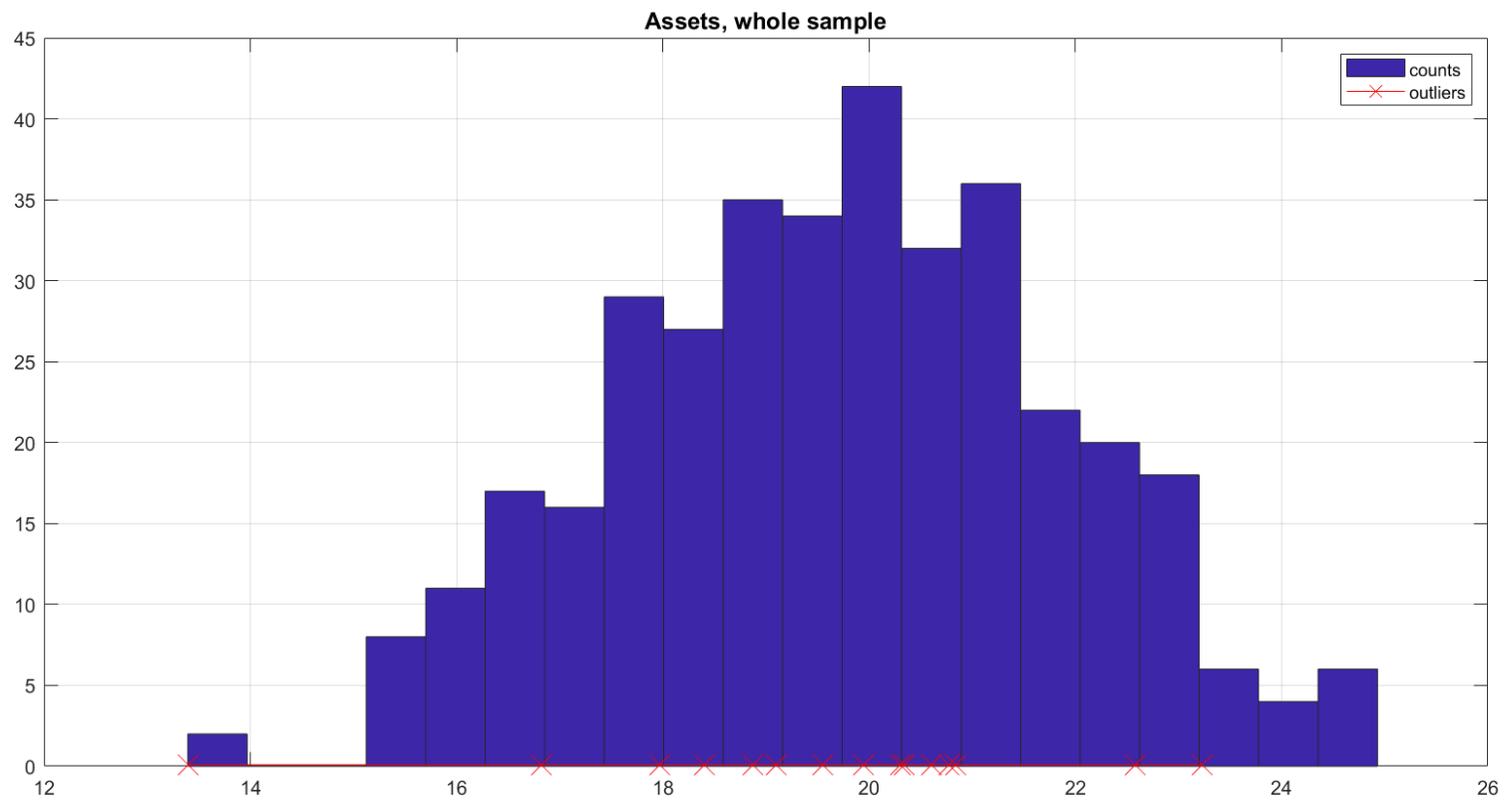

Figure 11.

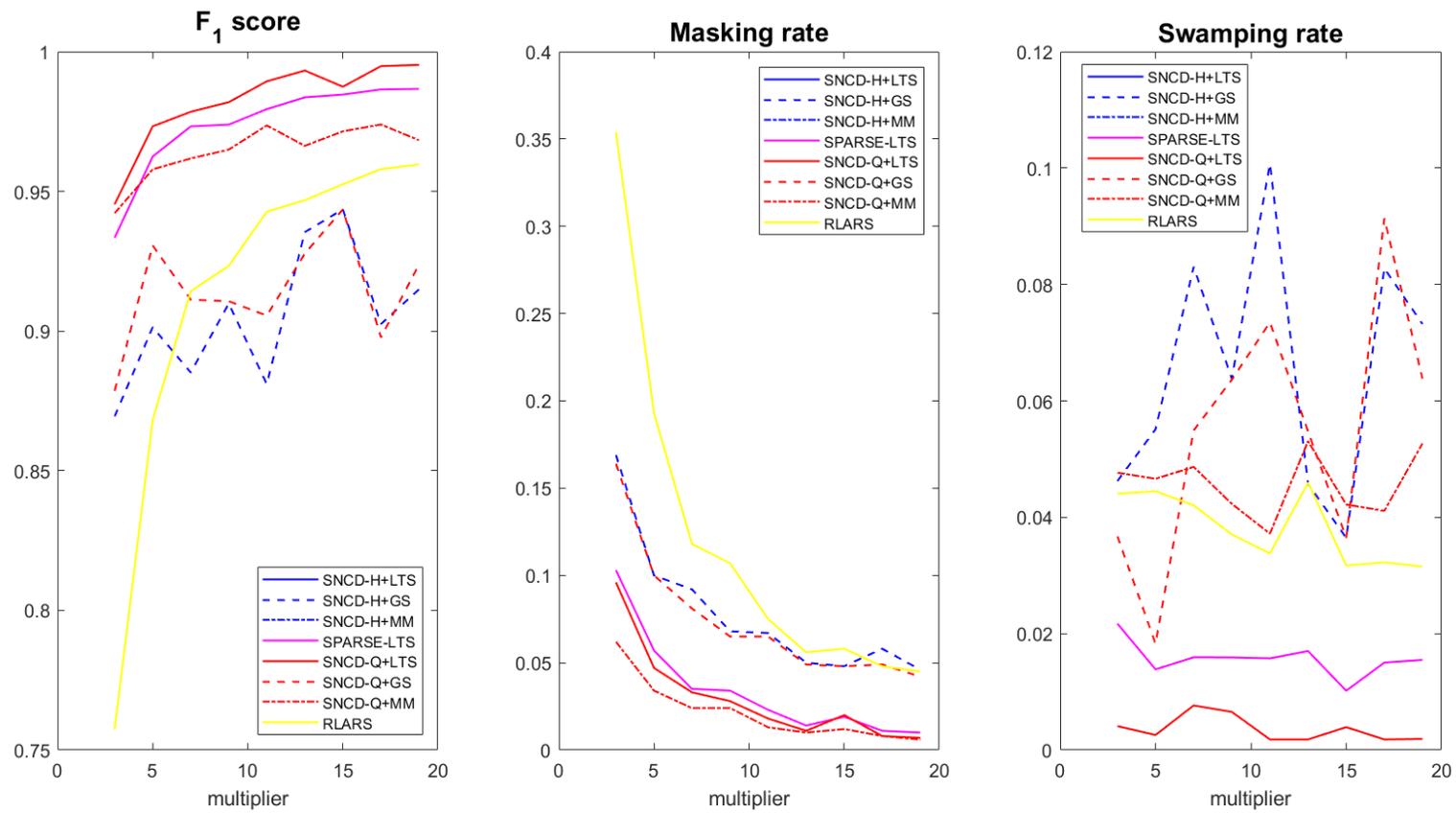

Figure A.1.

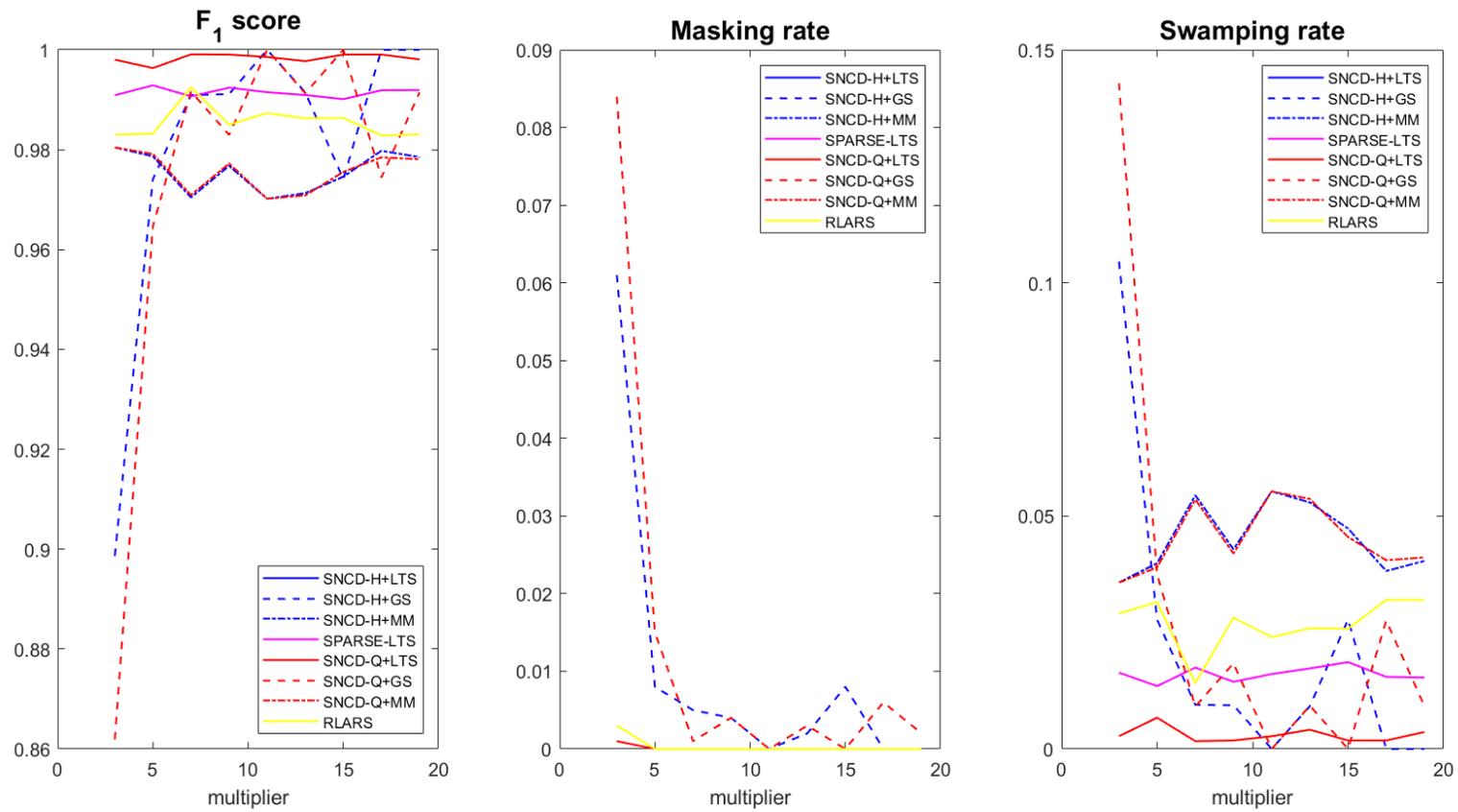

Figure A.2.

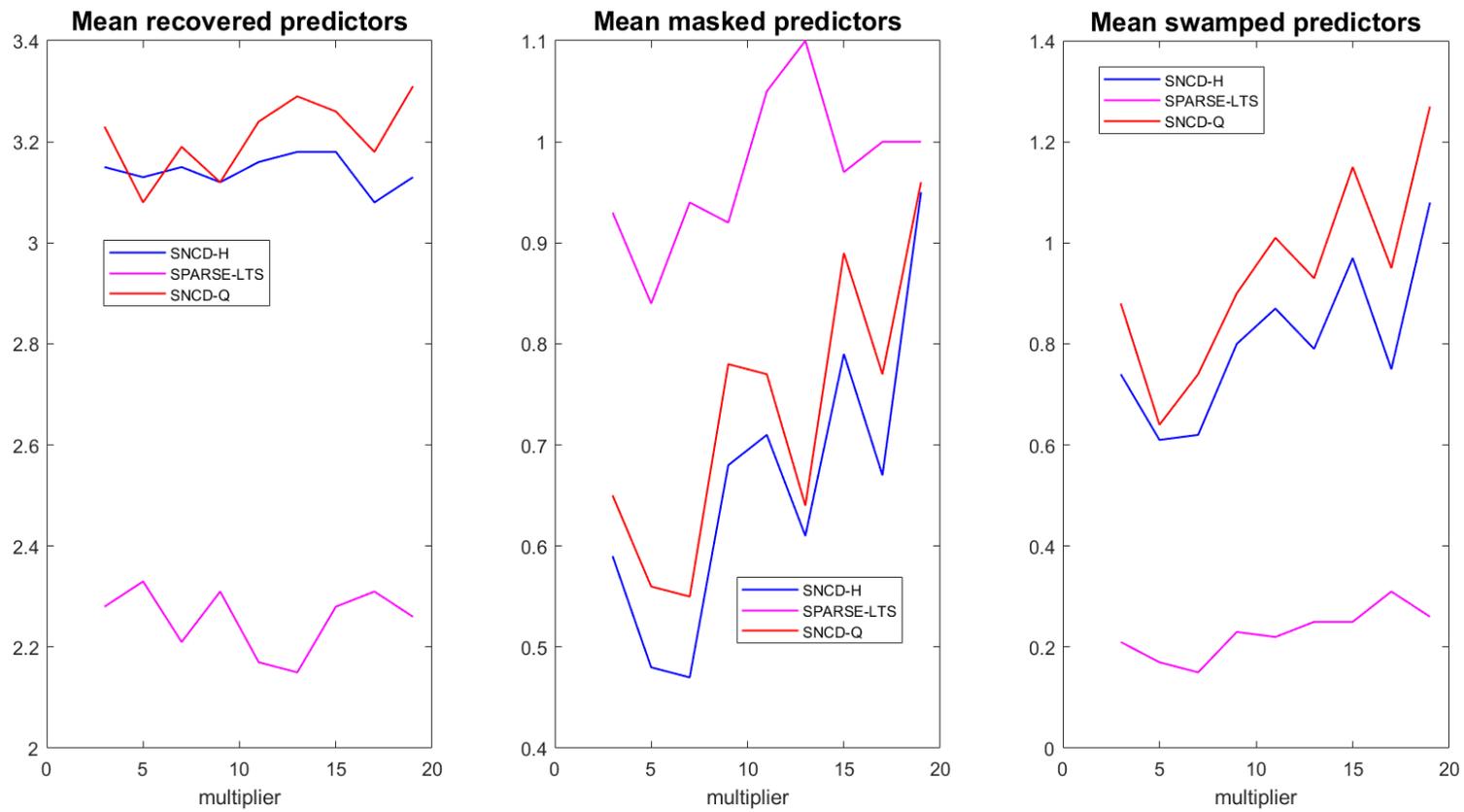

Figure A.3.

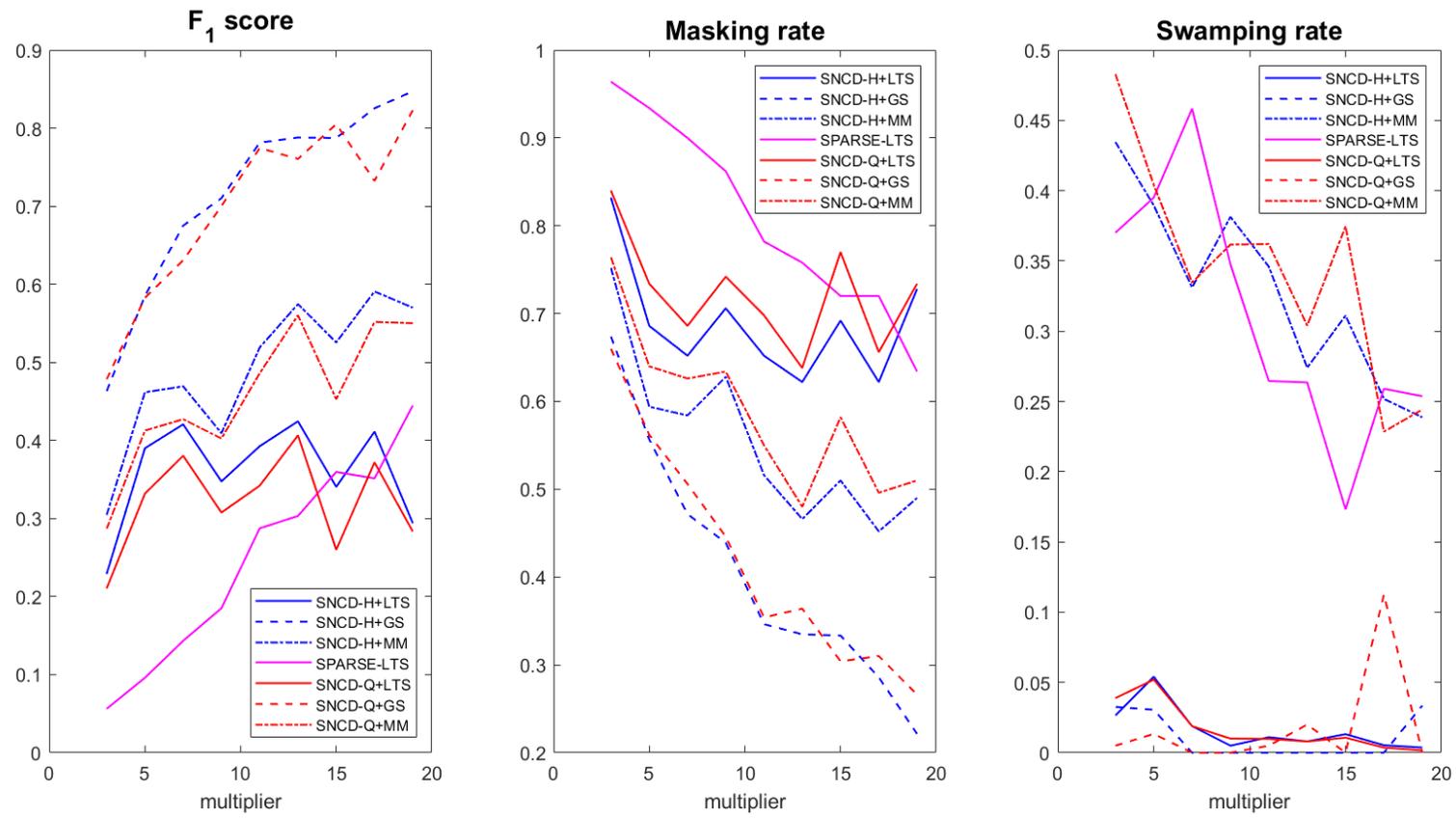

Figure A.4.

**List of figure captions**

Figure 1. Scenario 1a: F-score, masking rate and swamping rate of SNCD-H+LTS, SNCD-H+GS, SNCD-H+MM, SNCD-Q+LTS, SNCD-Q+GS, SNCD-Q+MM, SPARSE-LTS and RLARS, for m = 3,5,⋯,19.

Figure 2. Scenario 1b: F-score, masking rate and swamping rate of SNCD-H+LTS, SNCD-H+GS, SNCD-H+MM, SNCD-Q+LTS, SNCD-Q+GS, SNCD-Q+MM, SPARSE-LTS and RLARS, for m = 3,5,⋯,19.

Figure 3. Scenario 1c: Mean recovered, masked and swamped predictors by SNCD-H, SNCD-Q, SPARSE-LTS and RLARS, for m = 3,⋯,19.

Figure 4. Scenario 1c: F-score, masking rate and swamping rate of SNCD-H+LTS, SNCD-H+GS, SNCD-H+MM, SNCD-Q+LTS, SNCD-Q+GS, SNCD-Q+MM, SPARSE-LTS and RLARS, for m = 3,⋯,19.

Figure 5. Scenario 2c: Mean recovered, masked and swamped predictors by SNCD-H, SNCD-Q, SPARSE-LTS and RLARS, for m = 3,5,⋯,19.

Figure 6. Scenario 2c: F-score, masking rate and swamping rate of SNCD-H+LTS, SNCD-H+GS, SNCD-H+MM, SNCD-Q+LTS, SNCD-Q+GS, SNCD-Q+MM, SPARSE-LTS and RLARS, for m = 3,5,⋯,19.

Figure 7: Scenario 3a: F-score, masking rate and swamping rate of SNCD-H+LTS, SNCD-H+GS, SNCD-H+MM, SNCD-Q+LTS, SNCD-Q+GS, SNCD-Q+MM, SPARSE-LTS and RLARS, for m = 3,⋯,19.

Figure 8. Scenario 3c: Mean recovered, masked and swamped predictors by SNCD-H, SNCD-Q, SPARSE-LTS and RLARS, for m = 3,5,⋯,19.

Figure 9. Scenario 3c: F-score, masking rate and swamping rate of SNCD-H+LTS, SNCD-H+GS, SNCD-H+MM, SNCD-Q+LTS, SNCD-Q+GS, SNCD-Q+MM, SPARSE-LTS and RLARS, for m = 3,5,··· ,19.

Figure 10. Estimated residuals' distribution. Recovered conditional outliers are marked by a red cross.

Figure 11. Banks' log-size distribution. Recovered conditional outliers are marked by a red cross.

Figure A.1 Scenario 6b: F-score, masking rate and swamping rate of SNCD-H+LTS, SNCD-H+GS, SNCD-H+MM, SNCD-Q+LTS, SNCD-Q+GS, SNCD-Q+MM, SPARSE-LTS and RLARS, for m = 3,5,··· ,19.

Figure A.2 Scenario 7b: F-score, masking rate and swamping rate of SNCD-H+LTS, SNCD-H+GS, SNCD-H+MM, SNCD-Q+LTS, SNCD-Q+GS, SNCD-Q+MM, SPARSE-LTS and RLARS, for m = 3,5,··· ,19.

Figure A.3. Scenario 5c: Mean recovered, masked and swamped predictors by SNCD-H, SNCD-Q, SPARSE-LTS, for m = 3,··· ,19.

Figure A.4. Scenario 5c: F-score, masking rate and swamping rate of SNCD-H+LTS, SNCD-H+GS, SNCD-H+MM, SNCD-Q+LTS, SNCD-Q+GS, SNCD-Q+MM, SPARSE-LTS and RLARS, for m = 3,5,··· ,19.